\documentclass[prx,a4paper,twocolumn,showpacs,floatfix,superscriptaddress,amsmath,preprintnumbers,citeautoscript,aps,longbibliography]{revtex4-2}
\pdfoutput=1
\usepackage{dcolumn,graphicx,color,booktabs,microtype,afterpage}
\makeatletter\renewcommand{\fnum@figure}[1]{\figurename~\thefigure.}\makeatother
\makeatletter\renewcommand{\fnum@table}[1]{\tablename~\thetable.}\makeatother
\usepackage[colorlinks,plainpages=false,linkcolor=blue,urlcolor=blue,citecolor=blue,pdfpagemode=UseNone,pdfstartview=FitBH]{hyperref}

\usepackage{bm}% bold math
\newcommand{\cas}{CeAgSb$_2$}

\newcolumntype{/}{D{/}{/}{2,2}}  % / as a delimiter
\newcolumntype{.}{D{.}{.}{0}}  % / as a delimiter

\begin{document}

\title{Magnetic field-induced softening of spin waves and hard-axis order in Kondo-lattice ferromagnet~CeAgSb$_{2}$}

\author{S. E. Nikitin}
\affiliation{Paul Scherrer Institute (PSI), CH-5232 Villigen, Switzerland}
\affiliation{Max-Planck-Institut f\"{u}r Chemische Physik fester Stoffe, D-01187 Dresden, Germany}
\author{A. Podlesnyak}
\affiliation{Neutron Scattering Division, Oak Ridge National Laboratory, Oak Ridge, TN 37831, USA}
\author{J. Xu}
\affiliation{Helmholtz-Zentrum Berlin f\"{u}r Materialien und Energie GmbH, Hahn-Meitner-Platz 1, D-14109 Berlin, Germany}
\affiliation{Heinz Maier-Leibnitz Zentrum (MLZ), Technische Universit\"{a}t M\"{u}nchen, 85748 Garching, Germany}
\author{D. Voneshen}
\affiliation{ISIS, STFC, Rutherford Appleton Laboratory, Chilton, Didcot OX11 0QX, UK}
\affiliation{Department of Physics, Royal Holloway University of London, Egham, Tw20 0EX, UK}
\author{Manh Duc Le}
\affiliation{ISIS, STFC, Rutherford Appleton Laboratory, Chilton, Didcot OX11 0QX, UK}
\author{S. L. Bud'ko}
\affiliation{Ames Laboratory U.S. DOE, Iowa State University, Ames, Iowa 50011, USA}
\affiliation{Department of Physics and Astronomy, Iowa State University, Ames, Iowa 50011, USA}
\author{P. C. Canfield}
\affiliation{Ames Laboratory U.S. DOE, Iowa State University, Ames, Iowa 50011, USA}
\affiliation{Department of Physics and Astronomy, Iowa State University, Ames, Iowa 50011, USA}
\author{D. A. Sokolov}
\affiliation{Max-Planck-Institut f\"{u}r Chemische Physik fester Stoffe, D-01187 Dresden, Germany}

\begin{abstract}
A significant number of Kondo-lattice ferromagnets order perpendicular to the easy magnetization axis dictated by the crystalline electric field. The nature of this phenomenon has attracted considerable attention, but remains poorly understood.
In the present paper we use inelastic neutron scattering supported by magnetization and specific heat measurements to study the spin dynamics in the hard-axis ferromagnet \cas.
In the zero field state we observed two sharp magnon modes, which are associated with Ce ordering and extended up to $\approx 3$~meV with a considerable spin gap of 0.6~meV.
Application of a magnetic field perpendicular to the moment direction reduces the spectral intensity and suppresses the gap and significantly enhances the low-temperature specific heat at a critical field of $B_{\mathrm{c}}\approx 2.8$~T via a mean-field-like transition.
Above the transition, in the field polarized state, the gap eventually reopens due to the Zeeman effect.
We modeled the observed dispersion using linear spin-wave theory (LSWT) taking into account the ground state $\Gamma_6$ doublet and exchange anisotropy. Our model correctly captures the essential features of the spin dynamics including magnetic dispersion, distribution of the spectral intensity as well as the field-induced behavior, although several minor features remain obscure.  The observed spectra do not show significant broadening due to the finite lifetime of the quasiparticles. Along with a moderate electronic specific heat coefficient $\gamma = 46$~mJ/mol.K$^2$ this indicates that the Kondo coupling is relatively weak and the Ce moments are well localized.
Altogether, our results provide profound insight into the spin dynamics of the hard-axis ferromagnet \cas\ and can be used as solid ground for studying magnetic interactions in isostructural compounds including CeAuSb$_2$, which exhibits nematicity and unusual mesoscale magnetic textures.
\end{abstract}

\maketitle

\section{Introduction}

Ferromagnets are the simplest systems showing spontaneous time-reversal symmetry breaking. In the ordered state, the magnetic moments of all individual ions are pointed along the same direction creating a net moment. Usually, the direction of the spontaneous magnetisation in ferromagnets is dictated by the magnetocrystalline anisotropy, which aligns the ordered moment along the preferred crystallographic direction due to a combination of the spin-orbit coupling (SOC) and the crystalline electric field (CEF). The CEF couples the lattice to the electronic orbital angular momentum and this in turn is coupled to the spins via the SOC. This generally leads to a single crystallographic direction, which the spins either prefer to be parallel (easy-axis anisotropy) or perpendicular (easy-plane anisotropy) to. In 4$f$ intermetallic compounds, the primary exchange interaction is the indirect Ruderman-Kittel-Kasuya-Yosida (RKKY) exchange, which is long-ranged and oscillatory~\cite{ruderman1954indirect, kasuya1956theory,yosida1957}. This results in a competition between antiferromagnetic and ferromagnetic exchange interactions, which can lead to frustrated magnetic structures~\cite{mattis1962role}. In reality the majority of Ce and Yb-based intermetallics are antiferromagnets. Only a small number of ferromagnetic 4$f$ intermetallics are known and most of them order along the hard-axis showing the Kondo-lattice behaviour~\cite{hafner19, ahamed2018rare, brando2016metallic}.

Tetragonal CeAgSb$_2$ (space group $P 4/nmm$, No. 129) is a rare example of a ferromagnetic Kondo-lattice material. The Kondo coupling between localised Ce ions and conduction electrons sets at a temperature near or slightly above the Curie temperature, $T_{\mathrm{C}}$ = 9.6~K, inferred from the moderately enhanced Sommerfeld coefficient $\gamma$ = 46~mJ/mole$\cdot$K$^2$ above $T_{\mathrm{C}}$ and a local maximum in the temperature dependence of the electrical resistivity~\cite{HOUSHIAR19951231,myers99,jobiliong2005,takeuchi2003}. In the tetragonal crystal structure, the CEF lifts the six-fold degeneracy of the Ce$^{3+}$ ground state multiplet and splits it into 3 Kramers doublets. Analysis of high-field magnetisation, thermal expansion, and specific heat measurements established the CEF scheme, which consists of a $\Gamma_{6}$ doublet ground state and two $\Gamma_{7}$ doublets at 5.9 and 12.5~meV with respect to the ground state~\cite{araki03}. An estimate of the leading terms to the CEF splitting parameters, specifically the B$^{0}_{2}$ term was obtained in Ref.~\cite{myers99}.

The ferromagnetism and a weak Kondo-lattice behaviour seem to co-exist at the lowest temperature as indicated by photoemission spectroscopy measurements~\cite{saitoh2016}. The ordered moment points along the $c$-axis only, and no antiferromagnetic component was found in previous neutron diffraction measurements~\cite{araki03}. Nevertheless, the ferromagnetism of CeAgSb$_{2}$ is quite remarkable, in a sense that the ordered moment points along the hard direction, and not along the $a$-axis although the high-temperature magnetic susceptibility is greater along the $a$-axis at $T>T_{\mathrm{C}}$~\cite{myers99,araki03}. It has been suggested, based on results of inelastic neutron scattering measurements, that the ordering along the hard-axis is due to strong exchange anisotropy, and exchange interaction, $J_z$, which couples $z$ components of the spin operator is much stronger compared to the in-plane exchanges $J_{xy}$~\cite{araki03}. An alternative explanation of this phenomenon involves strong anisotropic quantum critical spin fluctuations, which destabilize the order in the $ab$-plane and promote the hard-axis ferromagnetism~\cite{krueger14, hafner19}. An even more intriguing scenario involves Kondo fluctuations, which promote hard-axis ordering as recently discussed in Ref.~\cite{kwasigroch2020magnetic}. Here we report extensive single crystal inelastic neutron scattering (INS) measurements on high-quality samples of CeAgSb$_{2}$ and consider these scenarios.

\section{Bulk measurements}\label{Sec:BulkMeasurements}
\begin{figure}
\includegraphics[scale=0.45]{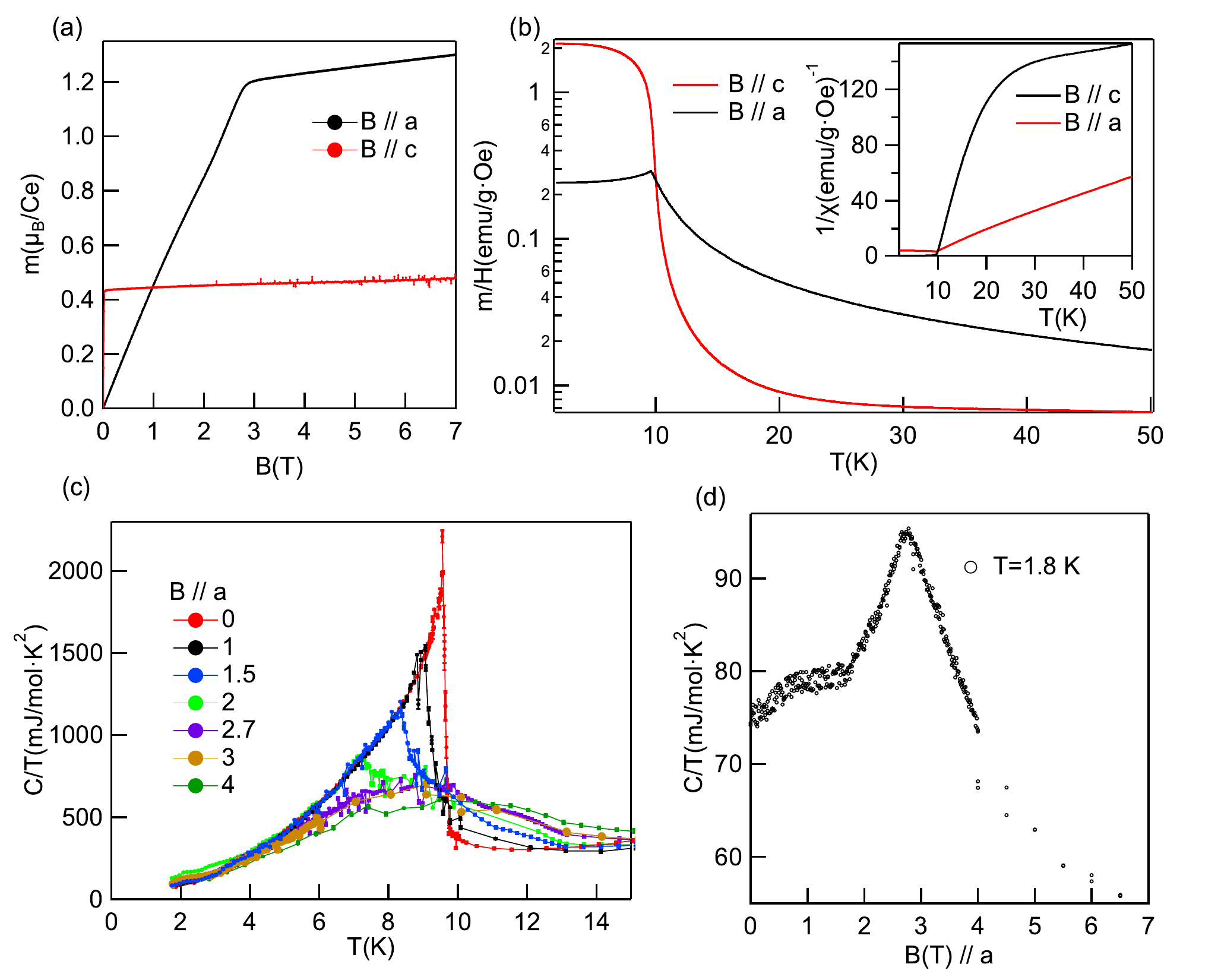}
\caption{(a)~Isothermal magnetisation $m$, measured for $B \parallel c$ and $B \parallel a$ axes at 1.8~K. (b) Temperature dependence of magnetisation measured for $B = 0.1~$T$ \parallel c$ and $B \parallel a$ axes. Note the crossing of magnetisations near $T_{\mathrm{C}} = 9.6$~K. Inset shows inverse magnetisation $1/m$ for both orientations of magnetic field. (c) Temperature dependence of the specific heat divided by temperature in magnetic fields applied along the $a$-axis. (d) Magnetic field dependence of the specific heat at 1.8 K with $B \parallel a$.}
\label{mvsh}
\end{figure}

As the first step we performed characterization of our samples by means of magnetization and specific heat measurements.
The magnetic field dependence of the magnetisation, $m(B)$, measured at 2~K in the geometry with $B \parallel c$ and $B \parallel a$ is shown in Fig.~\ref{mvsh}. Magnetization with $B \parallel c$ saturates at $\sim$0.47~$\mu_{\mathrm{B}}$/Ce ($\mu_{\mathrm{B}}$ is the Bohr magneton) below 0.01~T and shows a weakly field dependent behaviour at field up to 7~T consistent with the ordered moment oriented along the $c$-axis. When the magnetic field is applied along the $a$-axis, the field dependence of the magnetisation is linear at fields up to the inflection point at $\sim$2.8~T, above which the magnetisation increases only weakly reaching $\sim$1.29~$\mu_{\mathrm{B}}$/Ce at 7~T in agreement with Refs.~\cite{myers99,araki03,jobiliong2005,HOUSHIAR19951231}. The higher magnitude of the moment attained when $B \parallel a$ is due to asymmetry of the ground-state wavefunction and is consistent with the ground state being $\Gamma_{6}$ doublet. We conclude that in fields above $\sim$2.8~T the ordered moment is fully rotated and directed along the $a$-axis. Temperature dependence of the magnetisation measured with $B \parallel a$ shows a cusp at $T_{\mathrm{C}} = 9.6$~K. Above $T_{\mathrm{C}}$, the magnetisation with $B \parallel a$ is larger than with $B \parallel c$, suggesting that the $a$-axis is the easy axis. However, at temperatures slightly above $T_{\mathrm{C}}$, the magnetization curves cross and \cas\ orders along the hard $c$-axis in good agreement with data shown in Ref.~\cite{myers99}.

The specific heat shows a lambda-like anomaly at $T_{\mathrm{C}} = 9.6$~K, which corresponds to a continuous transition into the ferromagnetic state. The electronic part of the specific heat above the transition temperature approximates to $\sim$~46 mJ/mole$\cdot K^{2}$. The entropy $S$, released at the transition amounts to approximately 0.9~$R$ln2, where $R$ is the gas constant. These observations suggest that the ferromagnetism develops from a moderately correlated paramagnetic state, the Kondo lattice effect is relatively weak and one can treat the material like a local moment system. When the magnetic field is applied along the $a$-axis the anomaly in the specific heat decreases in size and shifts to low temperatures at fields up to 2~T. A putative quantum critical point was suggested at $\approx$2.8~T based on magnetization measurements~\cite{logg2013,kawasaki2018,jobiliong2005}. A broad feature, which develops at fields above 2~T most likely corresponds to a Zeeman's splitting of the ground state doublet. Specific heat measured as a function of the magnetic field  $B \parallel a$ at 1.8 K shows a sharp peak at 2.8~T, which corresponds to a full rotation of the ordered moment towards $a$-axis.

%%%%%%%%%%%%%%%%%%%%%%%%%%%%

%\section{Inelastic neutron scattering in field parallel to c-axis}

\section{Spin dynamics at zero field}\label{Sec:INSZeroField}

\subsection{CNCS data}

\begin{figure}[bth]
\includegraphics[width=1\linewidth]{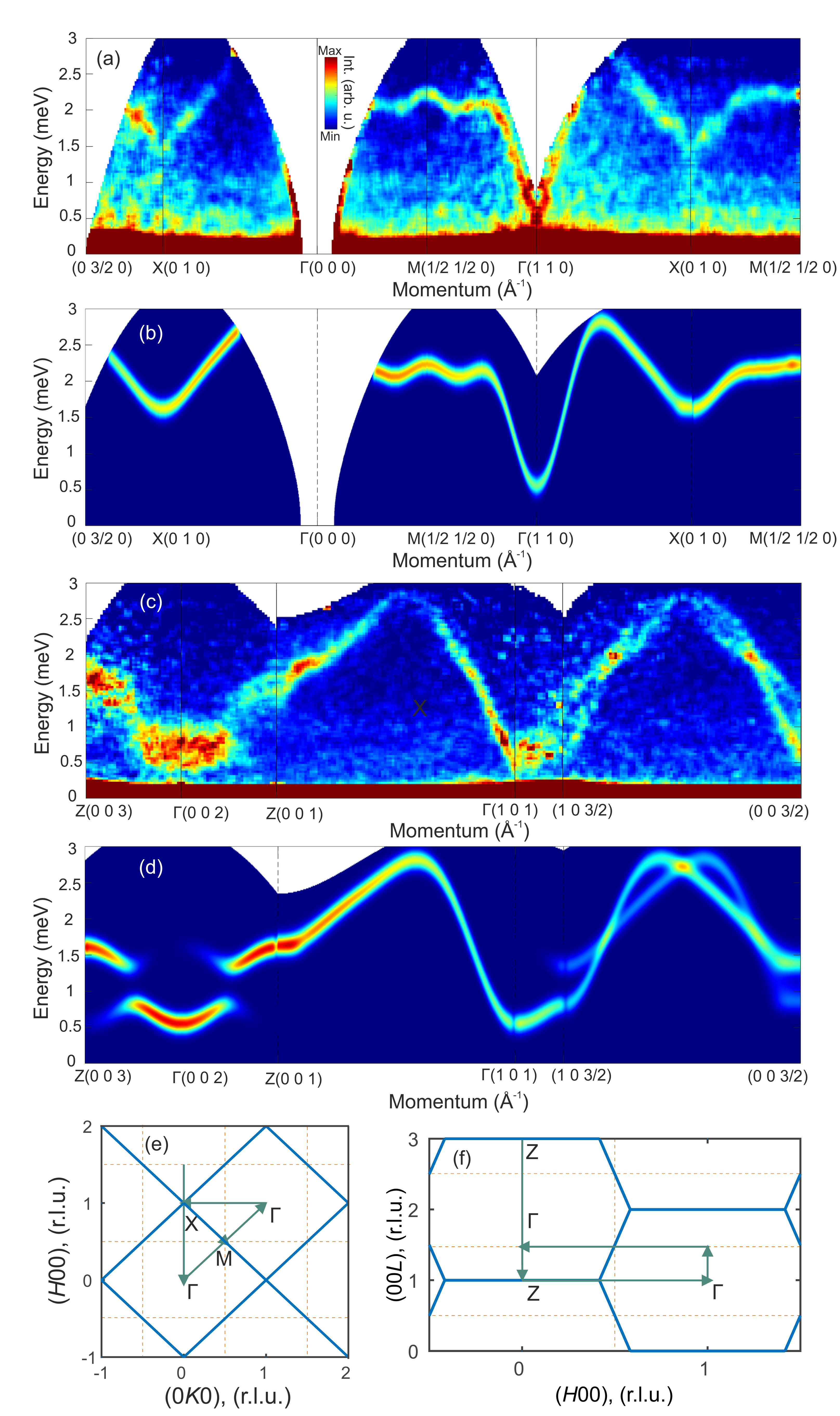}
\caption{~Spin-wave dispersion of CeAgSb$_{2}$ in ferromagnetic state at 1.7~K and 0~T measured at CNCS in $(HK0)$ (a) and $(H0L)$ (c) scattering planes. The data are integrated by $\pm 0.08$~\AA$^{-1}$ in two orthogonal directions. (b,d) Magnon spectra for the same path as in (a) and (d) panels, respectively, calculated using \textsc{SpinW} software. (e,f) Sketches of structural ``small'' and magnetic ``large'' Brillouin zones for $(H0L)$ and $(0KL)$ scattering planes shown by orange dotted and blue solid lines respectively. The green arrows show the paths in the reciprocal space for the spaghetti plot in (a-b) and (c-d) panels, respectively.}
\label{Fig:ZeroFieldMagnons}
\end{figure}

We start presentation of our INS data with the spectra collected at CNCS instrument at zero field deep inside the FM phase at 1.7~K, which are shown in Fig.~\ref{Fig:ZeroFieldMagnons} (a, c).
The excitation spectra consist of sharp spin-waves with a bandwidth of $\sim 3$~meV and a gap of $\sim 0.6$~meV.
Comparing different paths in Figs.~\ref{Fig:ZeroFieldMagnons} (a,c) one can clearly see that the dispersion along the $L$-direction is weaker compare to in-plane directions.
Such a behavior unambiguously indicates quasi-2D character of magnetic exchange interactions in the system, in agreement with naive expectations for the layered crystal structure of \cas.
It is worth noting that the out-of-plane dispersion branches (such as $Z-\Gamma-Z$ path in Fig.~\ref{Fig:ZeroFieldMagnons}(c)) are much broader than the in-plane ones, which are sharp and resolution-limited.
However, this broadening is not due to the intrinsic short lifetime of the magnons, but rather a consequence of the very steep dispersion in the orthogonal directions and the finite integration width ($\Delta{}Q = \pm0.08$~\AA$^{-1}$), which we have to apply to our data in order to achieve a reasonable signal-to-noise ratio in the spaghetti plots, see appendix~\ref{Sec:Fat_magnon} for details.

\begin{figure}[tb]
\includegraphics[width=.8\linewidth]{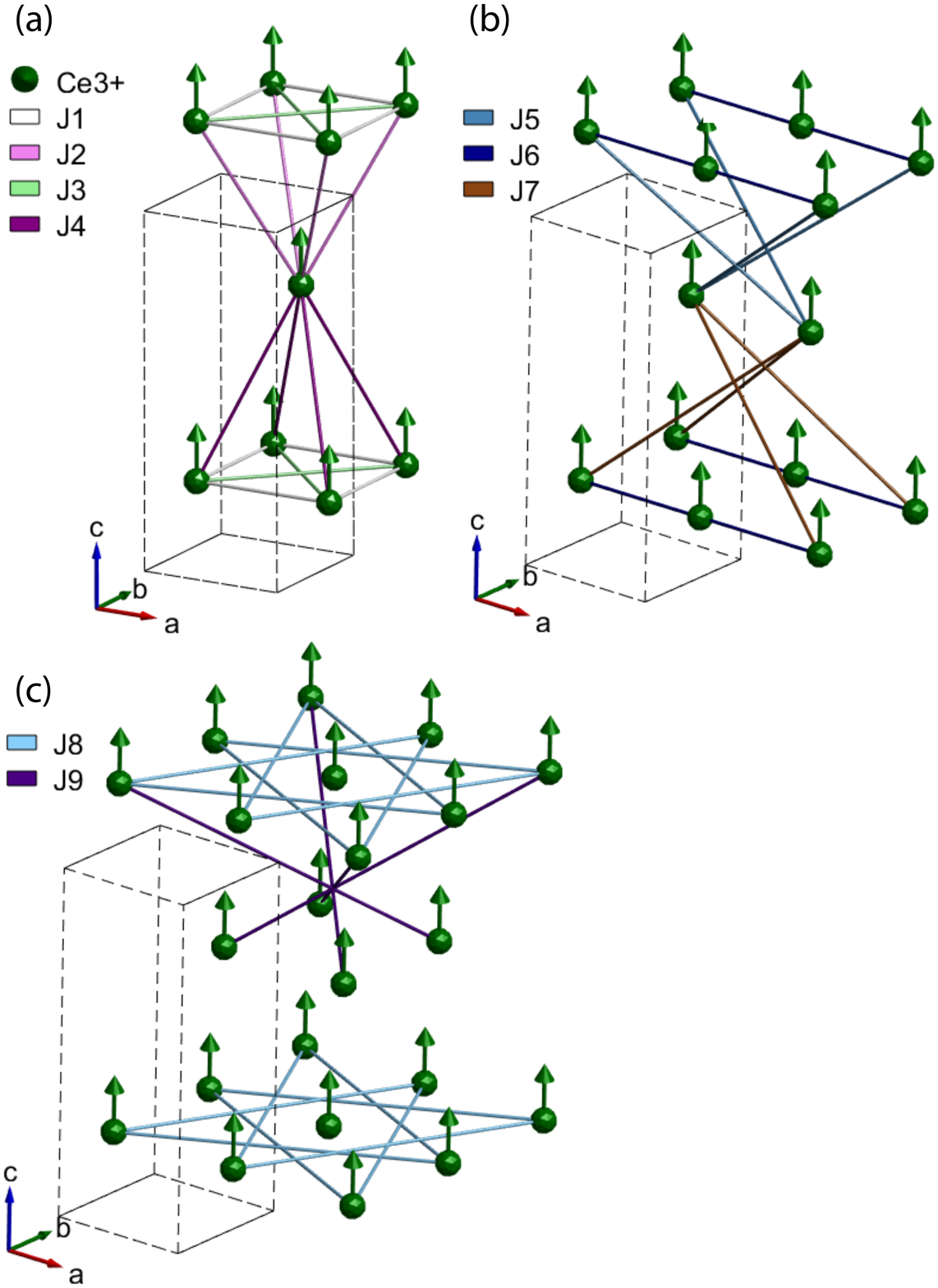}
\caption{~Zero-field magnetic structure and anisotropic exchange interactions in \cas\ in order of increasing Ce-Ce bond length. Only Ce ions are shown. (a) $J_1$ and $J_3$ are in-plane interactions along [100] and [110] directions, $J_2$ and $J_4$ are non-equivalent out-of-plane interactions due to the fact that $z$-coordinate of Ce $0.2388 \neq \frac{1}{4}$. (b) $J_5$, $J_7$, and $J_6$ are out of plane and in-plane interactions between Ce in neighboring unit cells. (c) $J_8$ and $J_9$ are interactions along [120] and [331] directions correspondingly.}
\label{Fig:structure}
\end{figure}

\begin{figure}[bth]
\includegraphics[width=1\linewidth]{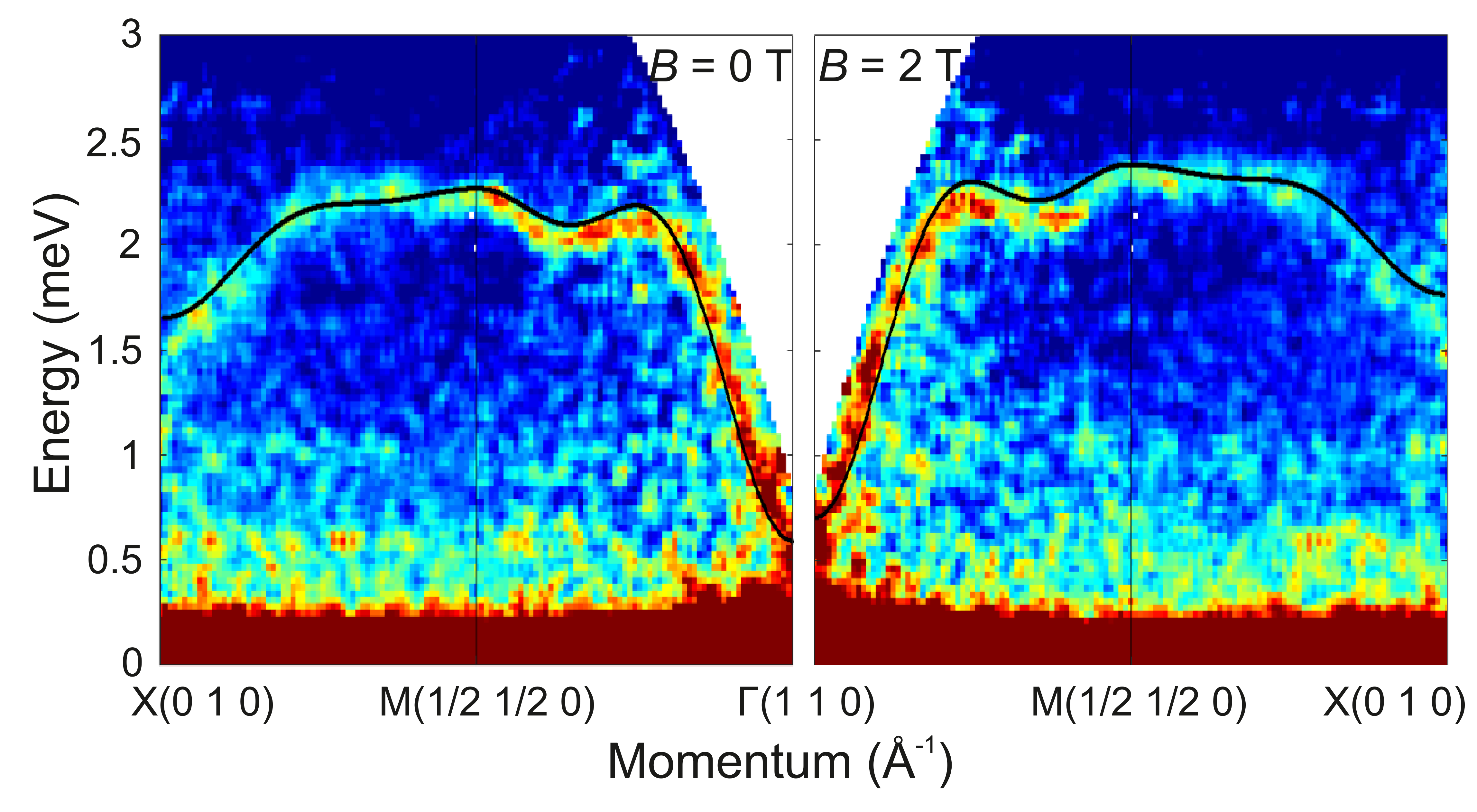}
\caption{~Spin-wave dispersion measured in $(HK0)$ scattering plane with magnetic field applied along the $c$-axis using CNCS spectrometer. Left and right part of the figure show the spaghetti plots along the same $\Gamma \rightarrow$ M $\rightarrow$ X path in the reciprocal space at 0 and 2~T as indicated in the panels. The data are integrated by $\pm 0.07$ (r.l.u.) in two orthogonal directions. Black lines shows results of spin-wave modeling. The width of the lines here and in all other figures represents intensities of the magnetic modes.
}
\label{Fig:CNCS_HK0}
\end{figure}

\begin{figure*}[tb]
\includegraphics[width=1\linewidth]{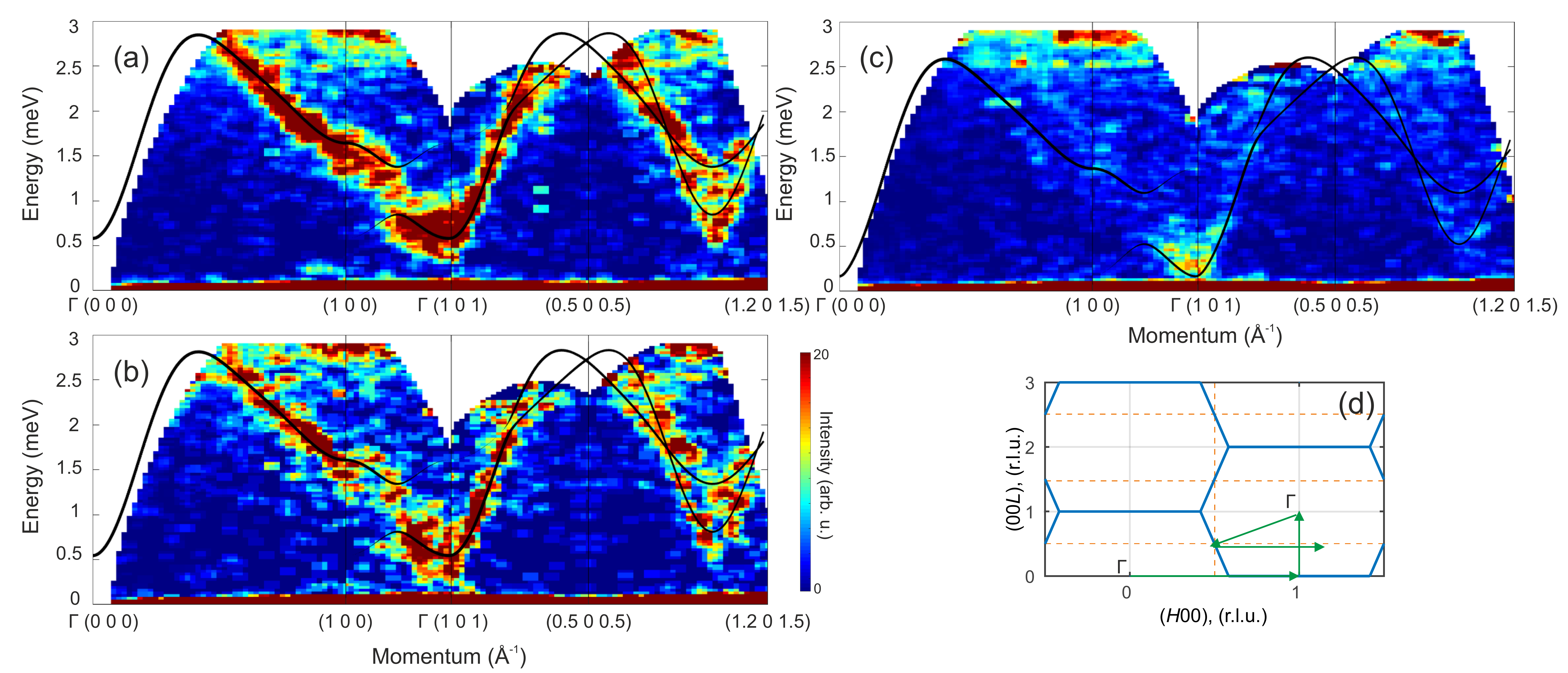}
\caption{~Spin wave dispersion measured in $(H0L)$ scattering plane with magnetic field applied along the $b$-axis using LET spectrometer. (a), (b) and (c) panels show the data collected at 0, 1 and 3~T. The data are integrated by $\pm 0.07$ (r.l.u.) in two orthogonal directions. The colorscale is identical for each panel. Black lines shows results of spin wave modeling.
(d) Orange and blue lines show sketches of the structural and magnetic Brillouin zones;} the green arrows show the path of spaghetti plots. 
\label{Fig:LETMagnons}
\end{figure*}

The unit cell of \cas\ contains two Ce ions, therefore one may expect to observe two magnon modes.
Our data indicate that in the majority of the Brillouin zone only a single mode carries spectral intensity. However, both modes are clearly visible in the middle of the magnetic Brillouin zone, when $L = \frac{2n+1}{2}$. As representative example see $(10\frac{3}{2}) \rightarrow (00\frac{3}{2})$ path of spaghetti plot displayed in Fig.~\ref{Fig:ZeroFieldMagnons} (c).
In addition, the two modes have a crossing point at zero field on the $(00\frac{3}{2}) \rightarrow (\frac{1}{2}0\frac{3}{2})$ path.

It is worth noting that in the previous report on spin dynamics in \cas\ the spectra were measured with significantly lower statistics and resolution using a triple-axis neutron spectrometer and the authors did not observe the second mode~\cite{araki03}. Therefore, for simplicity they considered a lattice with higher symmetry assuming Ce atomic coordinate $z = \frac{1}{4}$, when describing the spin dynamics. In contrast, our data clearly indicate a need to take into account the real crystal structure.
We also note that both magnetic and crystal structure of \cas\ have the same periodicity, and therefore the magnetic dispersion can be uniquely identified within a single crystallographic Brillouin zone. However, because \cas\ contains two Ce sites per unit cell, the intensity modulation follows the extended (also known as ``unfolded''~\cite{portnichenko2016magnon, tymoshenko2017pseudo}) Brillouin zone. The ``small'' crystallographic and the ``large'' magnetic Brillouin zones are schematically shown in Figs.~\ref{Fig:ZeroFieldMagnons} (e-f) for $(HK0)$ and $(H0L)$ scattering planes respectively and the high-symmetry points are marked according to the unfolded Brillouin zone.

\subsection{Linear spin-wave model}\label{Sec:INSZeroFieldLSWT}
\begin{figure}[tb]
\includegraphics[width=1\linewidth]{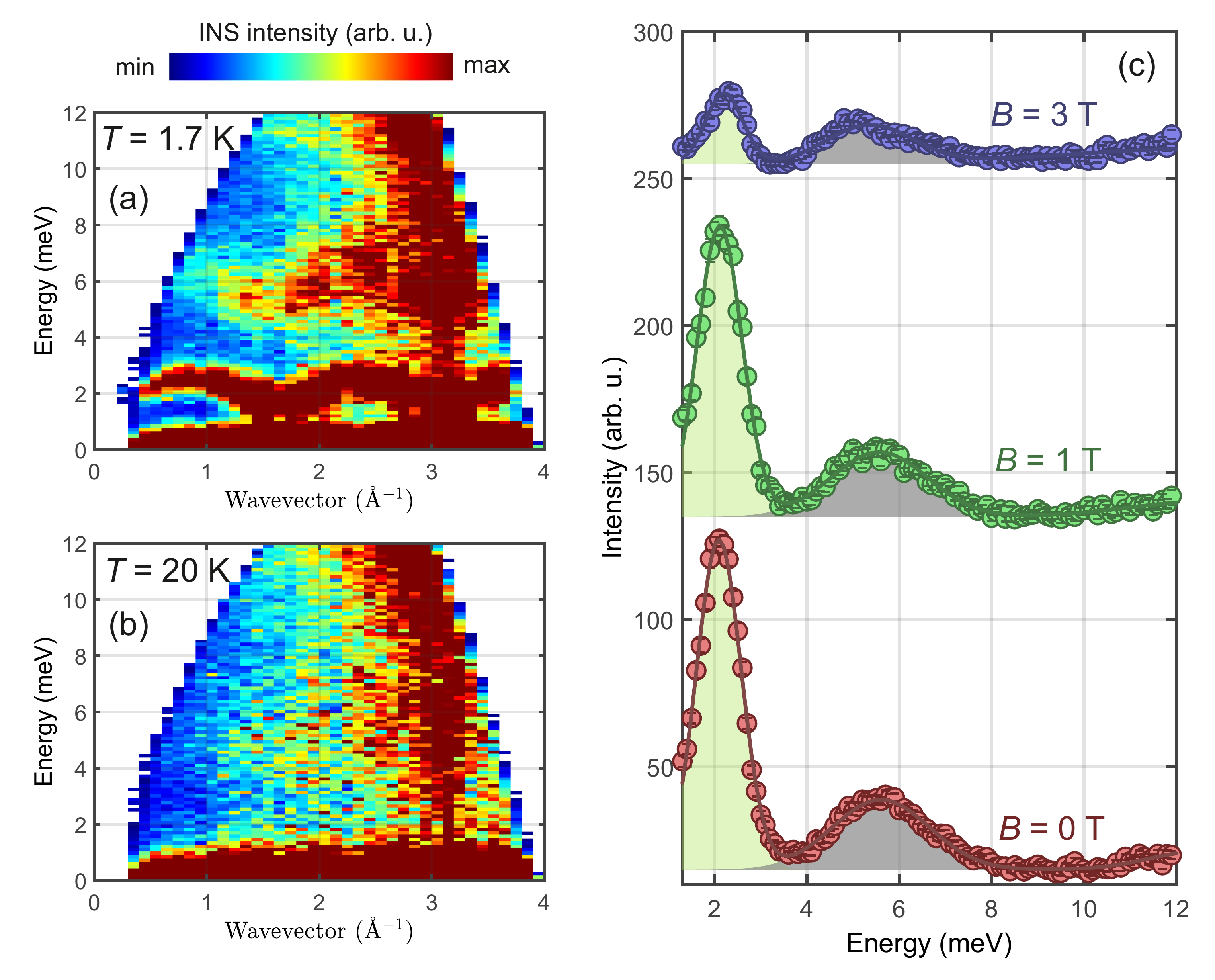}
\caption{~CEF excitations extracted from high-energy INS data measured using LET spectrometer with $E_{\mathrm{i}} = 14.9$~meV.
(a, b)~Powder-averaged INS signals of \cas\ measured at $B = 0$~T and different temperatures, as indicated in panels. Note that these patterns differ from the real powder signal, because the integration was performed within the $(H0L)$ scattering plane, rather than full reciprocal space.
Bright vertical excitation at high-$|\mathbf{Q}|$ is due to acoustic phonons emanating from $(200)$ structural peak.
(c)~Field dependence of magnetic excitations at $T = 1.7$~K and $B = 0, 1, 3$~T. The curves were obtained by integration of the INS signal at low $\mathbf{Q}$ within $H = 0\pm-1.5, K = 0\pm0.1; L = 0\pm1.5$~r.l.u.
Two shaded peaks at $\approx 2$ and $\approx 6$~meV show magnon and CEF excitations, respectively.
The data are  shifted vertically for clarity.
}
\label{Fig:CEF}
\end{figure}

The magnetic single-ion ground state of Ce$^{3+}$ ions in \cas\ is $\Gamma_6$ doublet~\cite{araki03}, while the first excited doublet is separated by substantial energy gap of $\sim6$~meV (see Fig.~\ref{Fig:CEF}), which is above the magnon bandwidth. Thus, in the first approximation, we can constrain ourselves and consider the low-lying doublet only. Therefore, we use pseudo-$S=1/2$ approximation, when describing the low-energy spin dynamics, as was done previously for many Yb- and Ce-based materials~\cite{Wu2016,paddison2017continuous, Nikitin2018, Wu2019, gao2019experimental}.

We model the low-energy spin dynamics using linear spin-wave theory (LSWT). In the general case, the spin Hamiltonian for a tetragonal system includes two main terms: (i) symmetric Heisenberg exchange interaction $J$, which controls the bandwidth and overall shape of the magnons; (ii) single-ion anisotropy $K_c(S^z)^2$, which defines the preferred orientation of the magnetic moments and opens a gap in the spectrum. However, in the case of $S = 1/2$ the single-ion anisotropy term is absent, and to reproduce the gap we used an exchange anisotropy instead, assuming $J^z \neq J^{xy}$. For simplicity we assumed that $\forall \{i,j\} \frac{J^z_{ij}}{J^{xy}_{ij}} =\mathrm{const}$. That is because the ratio $\frac{J^z_{ij}}{J^{xy}_{ij}}$ affects mainly the gap, and, therefore, an individual fitting of this ratio for each $J_{ij}$ increases drastically the number of free parameters and therefore is hardly feasible using our data.

Our model Hamiltonian is given by:
\begin{eqnarray}
 \mathcal{H} = \sum_{\langle i,j \rangle} && (J^z_{ij}{S^z_i S^z_{j}} + J^{xy}_{ij}{(S^x_i S^x_{j} + S^y_i S_{j}^y)}) +\nonumber \\
 &&\mu_0\hat{g}\mathbf{H}\sum_{i } {\mathbf{S}_i}.
 \label{Eq:SpinHamoltonian}
\end{eqnarray}
Summation in the first term runs over the different pairs of Ce ions as schematically shown in Fig.~\ref{Fig:structure}, and the second term shows the effect of external magnetic field, as will be discussed below. The $\hat{g}$-factor is a $3\times3$ diagonal matrix with $g_{x} = g_{y} = 3; g_{z} = 1$ for the $\Gamma_6$ doublet.

The crystal structure of \cas\ consists of square lattices of Ce ions, which are stacked along the $c$-axis, with a relative shift of $[\frac{1}{2}\frac{1}{2}0]$ with respect to each other (see Fig.~\ref{Fig:structure}). Moreover, because the atomic $z$-coordinate of Ce ions $0.2388 \neq \frac{1}{4}$, the distances between the layers are not equal to each other. Thus we can consider the crystal structure as a quasi-bilayer. The minimal set of the exchange interactions, which is required to couple all the Ce ions, consists of one in-plane exchange $J_1$ and two out-of-plane interactions within and between the bilayers $J_2$ and $J_4$.

\begin{figure}[tb]
\includegraphics[width=1\linewidth]{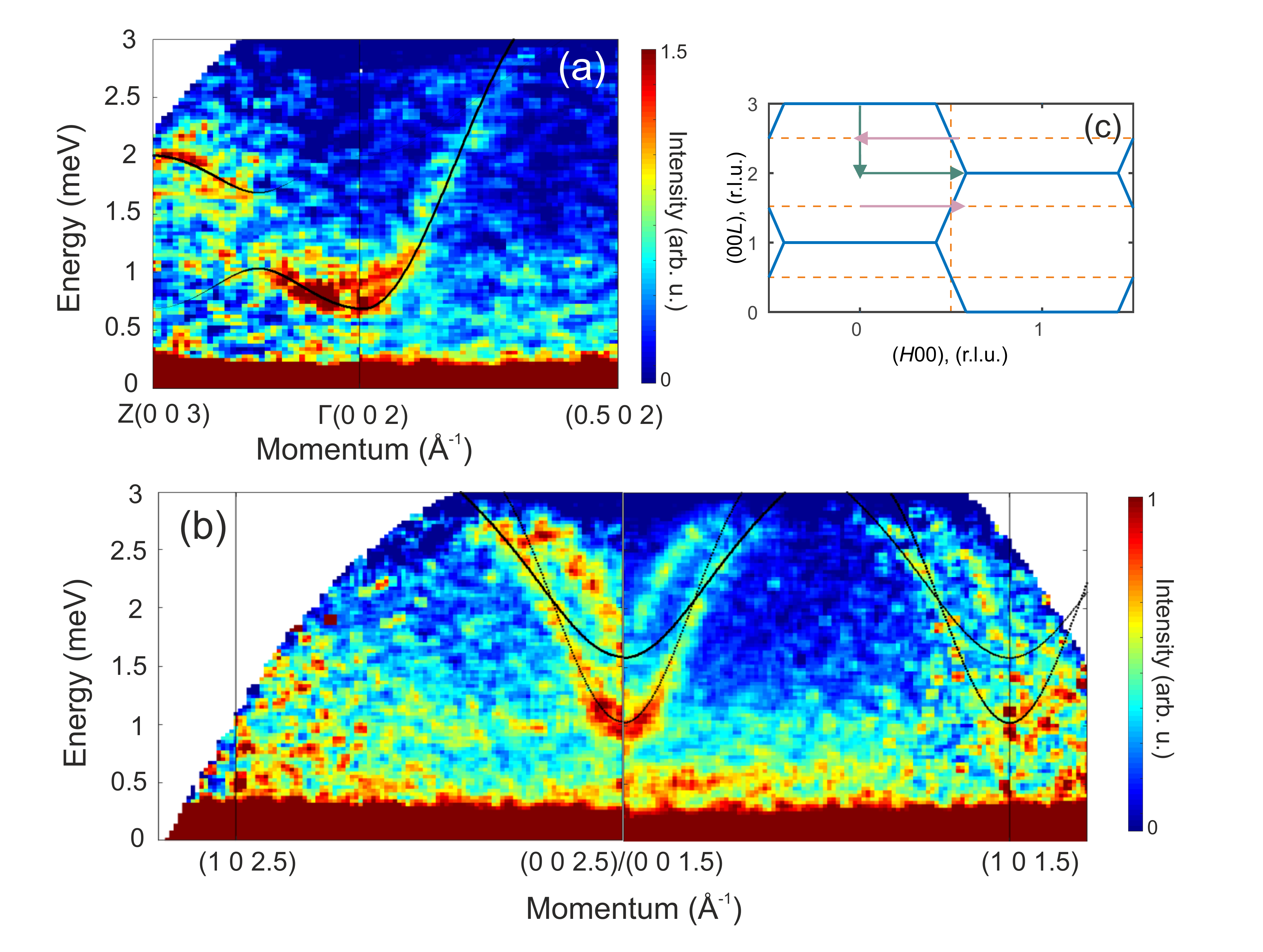}
\caption{~Spin wave dispersion measured in $(H0L)$ scattering plane with magnetic field $B = 6$~T applied along the $b$-axis using CNCS spectrometer. The data are integrated by $\pm 0.07$ (r.l.u.) in two orthogonal directions. Black lines shows results of spin wave modeling. (a) Dispersion along $(0 0 3) \rightarrow (0 0 2) \rightarrow (\frac{1}{2} 0 2)$ direction.
(b) Two energy-momentum slices along the equivalent directions of the Brillouin zone $(1 0 \frac{5}{2}) \rightarrow (0 0 \frac{5}{2})$ and $(0 0 \frac{3}{2}) \rightarrow (1 0 \frac{3}{2})$ clearly shows the absence of the mode crossing.
(c) Orange and blue lines show sketches of the structural and magnetic Brillouin zones; the green and purple arrows show the paths for panels (a) and (b) respectively.}
\label{Fig:6TMagnons}
\end{figure}

As the first step we fitted these parameters in order to describe the ground state magnetic structure and an overall amplitude of the magnons. Next, we begin to include further neighbor interactions to describe some specific features of the observed spectrum. The typical exchanges, which have clear manifestations is the long-distance coupling $J_8$ (the in-plane coupling along [120] direction), which is required to describe a sharp upturn at $M$ point of the Brillouin zone, see Fig.~\ref{Fig:ZeroFieldMagnons} (a). The ninth exchange interaction $J_9$ allowed us to qualitatively reproduce the specific shape of the dispersion on the $X \rightarrow M$ path (Fig.~\ref{Fig:ZeroFieldMagnons} (a)).

In such a way we computationally found a minimal set of parameters, which allowed us to qualitatively account for most features of the observed spectrum. Then we fitted our model to the energy of magnon modes at different $\mathbf{Q}$-points, which were determined by individual fitting of the constant-$\mathbf{Q}$ cuts at 109 points of reciprocal space to adjust all the exchange parameters. The uncertainties of exchange interactions were estimated by comparing results of the fittings with different initial parameters.

The neutron scattering spectra were calculated using a standard equation for the non-polarized inelastic neutron scattering cross-section as implemented in \textsc{SpinW} software~\cite{toth15, Squires_book_1978, lovesey1984theory, ZaliznyakLee_MNSChapter}:
\begin{align}
&\frac{d\sigma}{d\Omega_{\mathrm{f}} dE_{\mathrm{f}}}  =  \left(\frac{\mathbf{k}_{\mathrm{f}}}{\mathbf{k}_{\mathrm{i}}}\right) F(|\mathbf{Q}|)^2
\sum_{\alpha, \beta} (\delta_{\alpha, \beta} -\mathbf{\hat{Q}}_{\alpha} \mathbf{\hat{Q}}_{\beta})\times \nonumber \\
&\int_{-\infty}^{\infty}\langle\mathbf{S}_{j\alpha}(0)\mathbf{S}_{j'\beta}(t)\rangle
\rangle
\langle e^{i\mathbf{Q}\mathbf{r}_j(0)}e^{i\mathbf{Q}\mathbf{r}_{j'}(t)}  \rangle
e^{-iEt/\hbar} dt. \label{Eq:INS_cross_section_1}
\end{align}
The calculated spectra are shown in Fig.~\ref{Fig:ZeroFieldMagnons}(c,d). One can see a very good agreement between the calculated and observed patterns, including not only the dispersion, but also the redistribution of the spectral intensity between the modes over reciprocal space.  The ratio between $z$ and $xy$ components of the exchange interactions was found to be $\frac{J^z_{ij}}{J^{xy}_{ij}} = 1.37$. The obtained parameters of our spin Hamiltonian~\eqref{Eq:SpinHamoltonian} are given in table~\ref{tab:Exhange}.

The exchange interactions deduced in our work can be directly compared with the set of parameters obtained in Ref.~\cite{araki03} after a proper rescaling.
In our approach we describe the spin dynamic using conventional $S = 1/2$ formalism, while Araki et al. considered the anisotropic $\Gamma6$ ground-state doublet explicitly. Thus, the exchange interactions from~\cite{araki03}, $\{ I_{xy} \}$ should be rescaled by $g_a^2$ in order to be compared with the set of $\{ J_{xy} \}$ given in Table~\ref{tab:Exhange}. The anisotropy ratio can be recalculated as $I_z/(g_a^2 I_{xy})$. We found very good agreement for primary in-plane exchange interactions: $J_1 = 0.15$ ($I_1 = 0.162$), $J_3 = 0.25$ ($I_2 = 0.279$), $J_8 = 0.07$ ($I_6 = 0.063$), as well as for the anisotropy ratio, $J_{z}/J_{xy} = 1.37$ ($I_{z}/(g_a^2 I_{xy} = 1.333)$). Our parameters labeled as $J_i$ and Araki's are labeled as $I_i$ and given in parenthesis in meV units. The out-of-plane interactions are in poorer agreement, likely because of different definitions of the lattice used in two works as described above.

We note that in order to further justify use of $S = 1/2$ model we calculated the low-energy spin dynamics taking into account CEF explicitly by using random phase approximation (RPA) and compared the spectra with our LSWT data, see Appendix~\ref{Sec:mcphase} for details. The dispersion curves calculated by both methods are almost identical validating the use of the $S = 1/2$ model.

\section{Field-induced behavior of spin waves}

\subsection{INS results}

To reveal the effect of magnetic field on the spin dynamics of \cas\ we performed series of INS experiments in two geometries applying magnetic field along $c$ and $a$ axes. For the $\mathbf{B} \parallel [001]$ geometry we used CNCS instrument. Figure~\ref{Fig:CNCS_HK0} shows the spectra collected at 0 and 2~T along $\Gamma \rightarrow M \rightarrow X$ directions. One can see that the magnetic field enhances the spin gap by $\sim 0.1$~meV, while the details of the dispersion in both spectra look identical. The results are in excellent agreement with the expectation for the ferromagnet, polarized by the magnetic field.
The observed field-induced shift of the excitation energy well agrees with the Zeeman energy $\Delta{}E = g_z\mu_{\mathrm{B}}S\cdot2$~T = 0.11~meV calculated for $g_z = 1$.

However, more complex behavior is expected when magnetic field is applied along the in-plane direction $\mathbf{B} \parallel [010]$. In Sec.~\ref{Sec:BulkMeasurements} we used magnetization and specific heat measurements and showed that our sample exhibits a spin-reorientation transition at $B = 2.8$~T. Thus, excitations are also expected to be influenced by the transition.

We performed three experiments to study effect of the in-plane magnetic field on the spin waves using LET, CNCS and FLEXX neutron spectrometers. In order to avoid the sample torque when measuring the spectra above the $B_{\mathrm{c}}$ we heated up the sample above the Curie temperature, applied the target magnetic field and then field-cooled (FC) the sample to the base temperature ($T \approx 1.7$~K).

Figure~\ref{Fig:LETMagnons} shows the INS spectra collected using LET spectrometer at 0, 1 and 3~T, i.e. below and very close to the critical field.
The zero-field spectrum shown in Fig.~\ref{Fig:LETMagnons} (a) agrees well with the results obtained on the CNCS spectrometer, and the spin-wave dispersion calculated using the set of parameters deduced in Sec.~\ref{Sec:INSZeroFieldLSWT} describes well the observed magnon dispersion.
Application of 1~T field (i.e. below the critical value) has two effects: the spectrum gradually shifts towards lower energies, which is accompanied by the decreasing of the spectral intensity. However, the general shape of the dispersion curve remains unaffected. Figure~\ref{Fig:LETMagnons} (c) shows the INS spectrum collected at 3~T, very close to the critical field. One can see that both the spin gap and magnon intensity are suppressed further down. Because of the intensity's suppression we were able to resolve clearly only a weak excitation close to the $\Gamma$ point. The intensity of the signal at other parts of the Brillouin zone was below the detection limit.

Taking advantage of repetition rate multiplication we collected the spectra with higher incident energies, and thus resolved the field dependence of the lower CEF level. Figures~\ref{Fig:CEF}(a, b) show powder-averaged signal collected below and above the $T_{\mathrm{C}}$ respectively~\footnote{We note that these slices were obtained by $|\mathbf{Q}|$-averaging the signal within the $(H0L)$ scattering plane rather than the full reciprocal space, and therefore somewhat differ from a real powder spectra.}.
One can see that the low-temperature spectrum at low-$|\mathbf{Q}|$ consists of dispersive magnon excitations with bandwidth of $\approx 2.8$~meV and first CEF level at $\approx 5.6$~meV. We note that the first CEF level also exhibits finite dispersion and its position agrees well with previous reports. Due to use of relatively low $E_{\mathrm{i}}$ we were not able to resolve the second CEF level, which was observed previously at $12.5$~meV~\cite{araki03}.

The magnetic field significantly reduces the intensity of both magnetic peaks and slightly shifts them towards lower energy~\footnote{We discuss a possible origin of the suppression of the spectral intensity in Sec.~\ref{Sec:Suppression}}. However, even at 3~T, the CEF peak is separated by a considerable energy gap of $\approx 5.5$~meV. Thus, the excited CEF states are not likely to be responsible for the spin-reorientation transition, in agreement with the CEF calculations presented in Ref.~\cite{takeuchi2003}, which show that the magnetic field below $\sim 50$~T cause only a minor splitting of the Kramers doublets, as compare with the CEF gap.

The spectrum collected well above the field-induced transition at 6~T was measured using CNCS instrument and is shown in Fig.~\ref{Fig:6TMagnons}.
The excitation's intensity is much lower than one observed in 0~T in agreement with the LET results. In Fig.~\ref{Fig:6TMagnons}(a,b) we plot essentially all the observed signal, while at other parts of reciprocal space its intensity was below our detection limit. It is clear that the spin gap eventually reopens above the critical field shifting the spectrum towards the higher energies.

Moreover, we were able to resolve clearly the spin waves close to the crossing point. Panel (b) of Fig.~\ref{Fig:6TMagnons} shows two energy-momentum slices along equivalent directions of magnetic Brillouin zone $(1 0 \frac{5}{2}) \rightarrow (0 0 \frac{5}{2})$ and $(0 0 \frac{3}{2}) \rightarrow (1 0 \frac{3}{2})$, where we found mode crossing in the zero-field data, see Fig.~\ref{Fig:ZeroFieldMagnons} (c). One can clearly see that at 6~T both modes shows similar $\mathbf{Q}$-dependence and avoid the crossing in contrast to our calculations of the in-field magnon dispersion.

\begin{figure}[tb]
\includegraphics[width=1\linewidth]{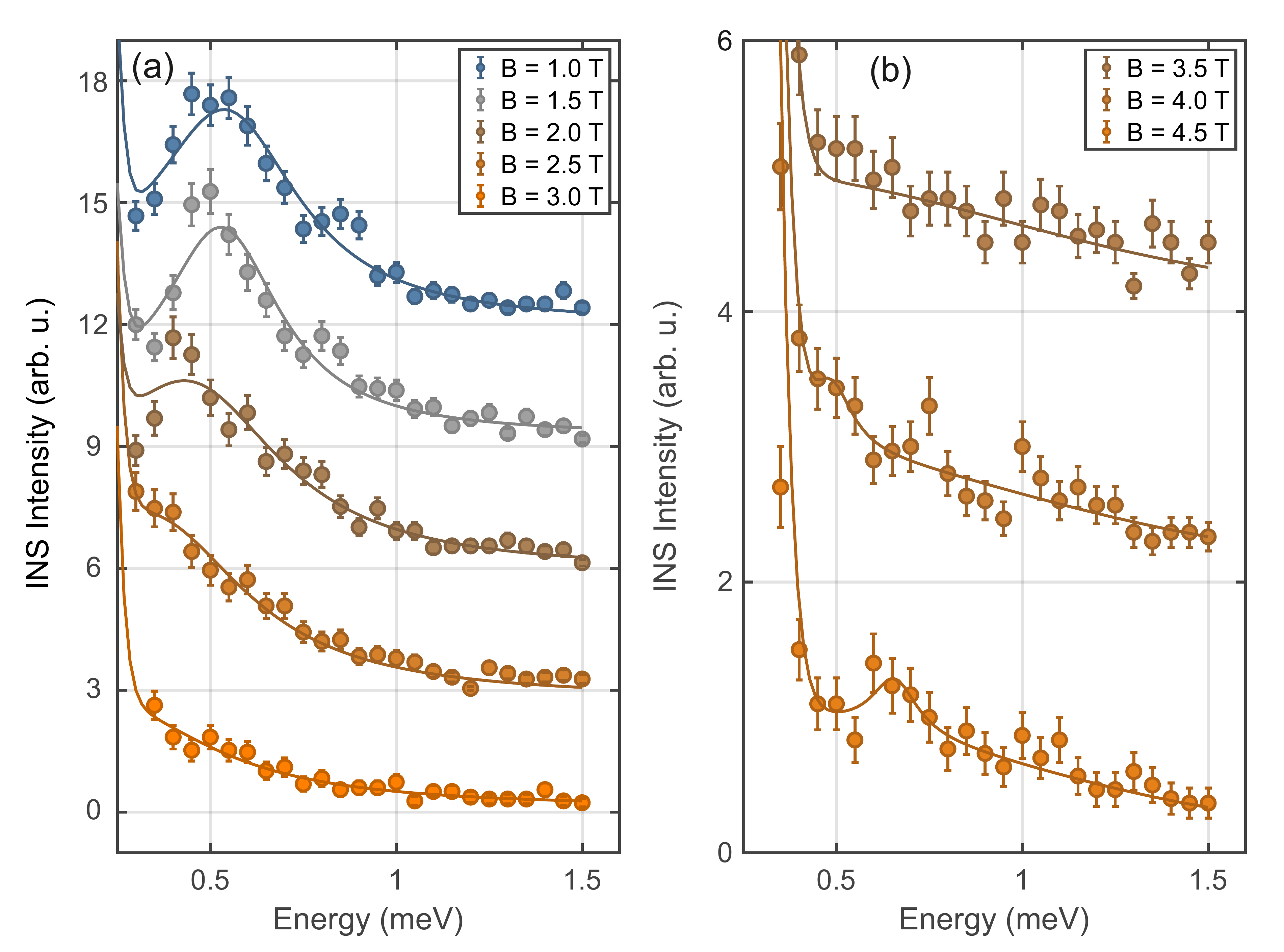}
\caption{INS spectra of \cas\ measured using FLEXX spectrometer at Gamma point $\mathbf{Q} = (101)$. Temperature was fixed to $T = 1.7$~K and magnetic fields is indicated in legends. Panels (a) and (b) show the data collected below and above the critical field $B_{\mathrm{c}}$ respectively. Data in both panels are vertically shifted for clarity. }
\label{Fig:FLEXX}
\end{figure}

Finally, we measured spin gap at $\Gamma$-point, $\mathbf{Q} = (101)$, with fine field step using FLEXX instrument. The measured curves are presented in Fig.~\ref{Fig:FLEXX}. Again, the signal intensity is suppressed with magnetic field and the peak position shifts towards lower energy. It fades below our resolution at 3~T and reappears at 4.5~T. We fitted the position of the peak and plotted it along with the values of spin gap obtained from CNCS and LET data in Fig.~\ref{Fig:SpinGap}.

%%%%%%%%%%%%%%%
\subsection{Modeling of field-induced spin dynamics}\label{Sec:InFieldModel}
The field-induced spin-reorientation transition is known to take a place in anisotropic ferromagnets, when the magnetic field is applied perpendicular to the direction of the ordered moment~\cite{pankrats2004magnetic, selter2020magnetic}. It originates from a competition between two energy scales: magnetic anisotropy and Zeeman field and our spin Hamiltonian~\eqref{Eq:SpinHamoltonian} also contains the key ingredients for such a behavior.

A prototypical example is the orthorhombic insulating ferromagnet PbMnBO$_4$~\cite{pankrats2016ferromagnetism}. In this material, application of a magnetic field perpendicular to the easy $a$-axis suppresses continuously the spin gap down to zero at the critical field, $B_{\mathrm{c}}$. With field increasing above $B_{\mathrm{c}}$, the gap eventually reopens and evolves almost linearly at higher fields. The field dependence of the spin gap for a tetragonal compound in the classical regime can be described by simple equations~\cite{pankrats2016ferromagnetism}:
\begin{eqnarray}
 &\hbar\omega(B) &= \hbar\omega_0\sqrt{1-\Big(\frac{B}{B_{\mathrm{c}}}\Big)^2}, \hspace{1cm}          B < B_{\mathrm{c}},  \label{Eq:field_dep1} \\
 &\hbar\omega(B) &= \hbar\omega_0\frac{B}{B_{\mathrm{c}}}\sqrt{1-\Big(\frac{B_{\mathrm{c}}}{B}\Big)}, \hspace{1cm}          B > B_{\mathrm{c}},  \label{Eq:field_dep2} \\
 &\hbar\omega_0 &= 2 \mu_{\mathrm{B}}gS B_{\mathrm{c}}, \label{Eq:field_dep3}
\end{eqnarray}
where $\hbar\omega_0 = 0.57(2)$~meV is the zero-field gap, which defines uniquely the critical field via Eq.~\eqref{Eq:field_dep3}, and is controlled by the anisotropy ratio $\frac{J_{z}}{J_{xy}}$. The $g_x = 3$ is known from the CEF calculations for $\Gamma_6$ doublet, while the critical field was extracted precisely from the thermodynamic measurements, $B_{\mathrm{c}} = 2.87(5)$~T.

\begin{figure}[tb]
\includegraphics[width=1\linewidth]{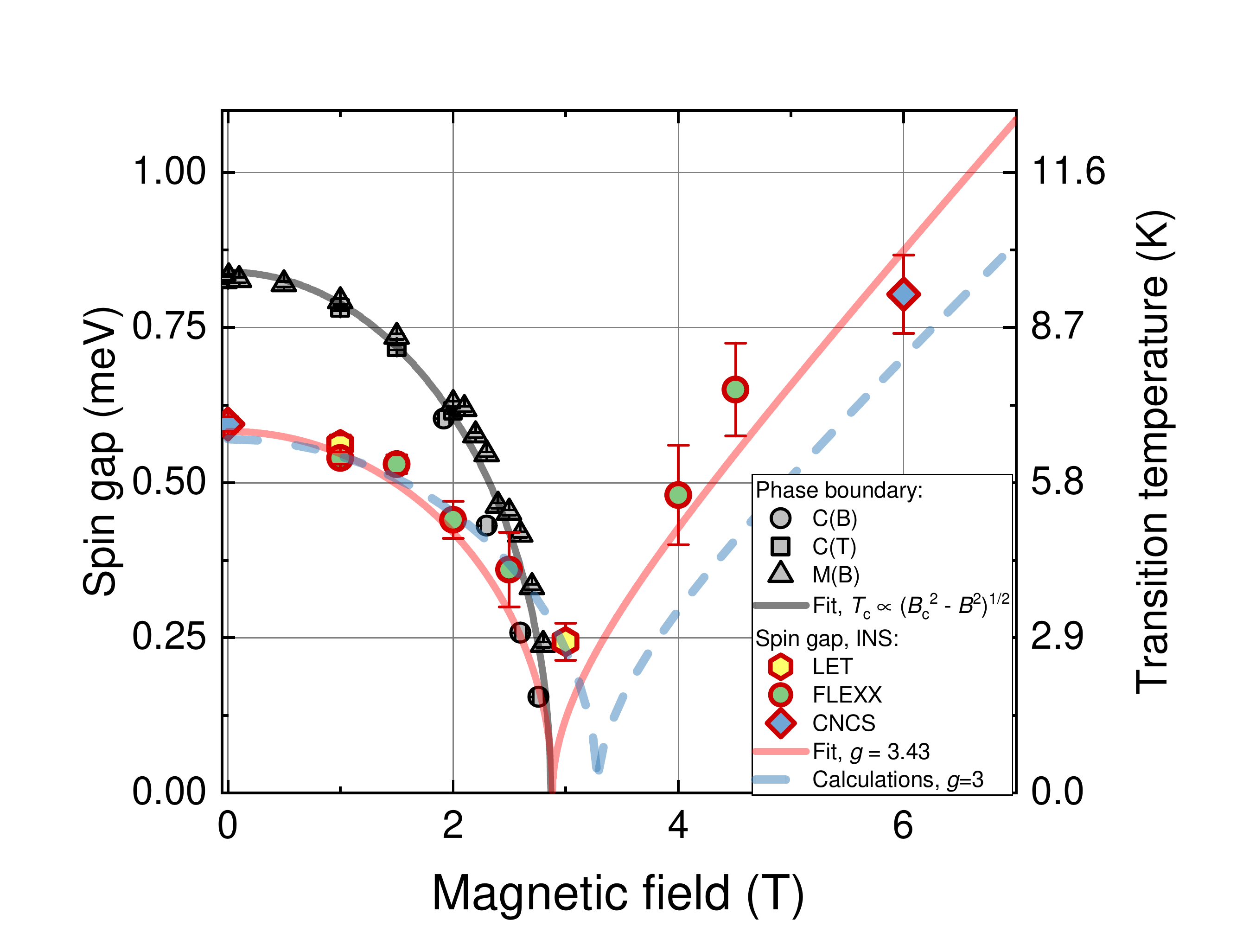}
\caption{~Spin gap at $\Gamma$-point calculated from the INS spectra (colored dots) and temperature of spin-reorientation transition deduced from thermodynamic measurements (grey dots) plotted as a function of magnetic field.
Red and blue lines show the spin gap calculated with Eqs.~\eqref{Eq:field_dep1}, \eqref{Eq:field_dep2} and \eqref{Eq:field_dep3} for $g = 3$ and $g = 3.43$. Black line is a fit of the transition temperature with $T_{\mathrm{c}} \propto \sqrt{B^2_{\mathrm{c}} - B^2}$.
The temperature and energy axes are shown on the same scale $E = k_{\mathrm{B}}T$.
}
\label{Fig:SpinGap}
\end{figure}

However we note that the Eq.~\eqref{Eq:field_dep3} with  $g_x = 3$ and $\hbar\omega_0 = 0.57(2)$ yields critical field  $B_{\mathrm{c}} =  3.28$~T above the experimentally determined value.
Having in mind that the field dependence of the gap and critical field were reliably determined from the experimental data we used Eqs.~\eqref{Eq:field_dep1}, \eqref{Eq:field_dep2}, \eqref{Eq:field_dep3} in order to fit the field dependence of the spin gap having $g$ factor as a free parameter starting from theoretically predicted $g_x = 3$, and the fitting yields a 15~\% increase of an effective $g_x-$factor, $g_x = 3.43(8)$.
The curves calculated for both values of $g$ are presented in Fig.~\ref{Fig:SpinGap}.
Some deviation between the calculated and experimental curve was observed at 3~T. This effect can be due to a small deviation of the magnetic field from the nominal value during the experiment along with the very large slope of $dE/dB$ in the vicinity of the critical field.

Finally, we discuss the origin of the disagreement between the measured and calculated magnon dispersion curves at $B = 6$~T, which are presented in Fig.~\ref{Fig:6TMagnons} (b).
At zero field, two magnon modes have a crossing point, and our spin-wave model correctly captures this behavior. However, the model predicts that application of magnetic field does not change the dispersion itself and in-field magnon dispersion can be rewritten as $\omega_n(\mathbf{q}, B) = \Delta(B) + \omega_n(\mathbf{q}, B = 0)$, where the spin gap, $\Delta(B)$, follows Eqs.~\eqref{Eq:field_dep1} and \eqref{Eq:field_dep2}.
However, the measured INS spectrum shown in Fig.~\ref{Fig:6TMagnons} (b) shows two branches, which do not cross at the predicted wavevector.
We speculate that it might have two different origins.
First of all, magnetostriction changes the inter-atomic distances, and consequently modifies the exchange interactions. However, the measured values of magnetostriction for $B = 6$~T applied along the $ab$ plane is only $4\times 10^{-6}$ to $4\times 10^{-5}$ depending on the direction~\cite{adroja2002thermal}. Thus, one would not expect significant modifications of the magnetic exchange parameters. On the other hand, the position of the mode crossing in our model is given by a delicate balance between out-of-plane couplings, and thus is sensitive to small variations of different exchanges.
The second explanation involves magnon-magnon interaction, which can cause a well-known anticrossing behavior~\cite{endoh1984resonant, kuroe2011hybridization, hayashida2015magnetic}. This can also explain the absence of mode crossing in our data, however, detailed analysis of magnon-magnon interaction goes beyond our simple linear spin-wave theory approximation and requires additional theoretical investigations.

\section{Discussion}

Recently, a generic mechanism, which explains why many metallic Kondo ferromagnets order along directions not favoured by the crystal field anisotropy was proposed in Ref.~\cite{krueger14}. CeAgSb$_{2}$ is well-positioned to test the predictions of this model. Specifically, based on bulk measurements, the material orders along the hard-axis at the temperature close to a Kondo temperature, the magnetic susceptibilities measured along the hard and easy directions cross slightly above $T_{\mathrm{C}} = 9.6$~K, and the ordered moment is rotated by application of a modest magnetic field from the hard axis to an easy plane, following simple predictions for classical magnets. Note that at the critical field the magnon gap is closed, which also causes significant enhancement of the low-temperature specific heat due to magnetic degrees of freedom, see Fig.~\ref{mvsh} (d). The theory argues that the low-energy particle-hole fluctuations can lead to the hard-axis ordering in the metallic ferromagnets at low temperatures due to the coupling to the electronic quantum fluctuations. The theory argue further that the switch of magnetic anisotropies on cooling and the eventual hard-axis ordering at the lowest temperatures in ferromagnetic YbNi$_{4}$P$_{2}$ is due to strong spin fluctuations transverse to the ordered moment. Our inelastic neutron scattering data show mostly the transverse fluctuations, which remain strong even at fields above the moment reorientation transition, which seems to rule out the fluctuations driven mechanism for the hard-axis ordering. We note that the Kondo coupling in CeAgSb$_{2}$ is probably weaker than in YbNi$_{4}$P$_{2}$ and the nature of the magnetic order is more conventional.

Our description of the low-energy magnetism of CeAgSb$_{2}$ is largely focused on anisotropic exchange interaction between local pseudo-$S = 1/2$ moments of Ce ions within the ground-state doublet. We used LSWT and were able to describe the zero-field magnon excitations quantitatively. These calculation however confirm that the exchange is stronger along the hard-axis by a factor of 1.37, and in the absence of competing phenomena such as magnetic frustration, stronger exchange usually dictates the direction of magnetic ordering. We further note, that if one takes into account the full multiplet, $J = 5/2$, it is necessary to rescale the obtained exchange parameters by corresponding $g$-factors, and thus the exchange anisotropy is increased to $J_z/J_{xy} = 12.4$, in the quantitative agreement with the value deduced in Ref.~\cite{araki03}. Our results clearly indicate that long-range exchange interactions, well beyond nearest neighbors, are essential. We further note that although the set of exchange interactions deduced in the presented work reproduces magnon dispersion at zero field and field-induced behavior of the spin gap, it is not defined uniquely and can be considered as set of $\textit{effective}$ interactions. This is because: (i) we took into account all interactions up to 9th nearest-neighbor providing a minimal set of interaction, which describe our data, however, the long-range nature of RKKY interaction implies that other exchange interactions can be also significant; (ii) $\frac{J^z_{ij}}{J^{xy}_{ij}}$ ratio can vary for each $\langle i, j \rangle$ pair, but we made a simplification by assuming $\forall \{i,j\} \frac{J^z_{ij}}{J^{xy}_{ij}} =\mathrm{const}$ to reduce the number of free parameters.

According to our results, the ground state of \cas\ is a simple, non-frustrated quasi-two dimensional FM with easy-axis exchange anisotropy. However, the RKKY interaction responsible for the magnetic ordering is rather sensitive to details of the electronic structure as well as inter-atomic distances, thus considerable variations of anisotropic exchange interactions are expected in RETX$_{2}$ series (RE= La, Ce, Pr, Nd, Sm; T = Cu, Ag, Au; X = Sb, Bi), which can stabilize various magnetic ground states characterized by different polarisation and propagation directions~\cite{myers99} or even to striped and multi-$\mathbf{Q}$ orders in CeAuSb$_{2}$~\cite{marcus2018} and similar compounds.

\section{Conclusion}
Summarizing, we have performed a comprehensive experimental study of the spin dynamics in \cas\ using inelastic neutron scattering.
The low-temperature excitation spectrum consists of a gapped magnon bands at low energy, which can be reasonably well described by a linear spin wave theory and a CEF excitation at $\approx 5.6$~meV, in agreement with previous reports.
Fitting of the magnon dispersion revealed a quasi-two-dimensional hierarchy of exchange interactions, with $|J_{ab}| > |J_c|$, which produce a non-frustrated FM ground state in \cas.
Application of in-plane magnetic field causes a spin-reorientation transition at 2.8~T, which is accompanied by a closing of the spin gap and increase of the low-temperature specific heat. Above $B_{\mathrm{c}}$ the gap eventually reopens due to the Zeeman effect. The overall behavior of the spin gap is well captured by theory predictions for a classical anisotropic FM.
We further found that two magnon modes do not cross in the field-polarized state, in contrast to the predictions of our model. We speculate that this behavior originates from a field-induced modification of the exchange interactions or magnon-magnon interactions, but further high-field INS experiments and advanced theory calculations are required to address this issue.

 \begin{table}[tb]
 \caption{~Exchange interactions, $J_{xy}$, in \cas\ determined by fitting of the 0~T dataset. All energies are given in (meV) units. Negative sign corresponds to FM exchange.} \label{tab:Exhange}
 \begin{ruledtabular}
 \begin{tabular}{ll| ll}
 Exchange & Value  & Exchange & Value \\
         \hline
    $J_1$ &  -0.15(3)  & $J_6$          &  0$\pm 0.02$ \\
    $J_2$ &  -0.04(4)  & $J_7$          & -0.014(6)    \\
    $J_3$ &  -0.25(3)  & $J_8$          & -0.070(2)    \\
    $J_4$ &  -0.02(4)  & $J_9$          & -0.076(1)     \\
    $J_5$ &  -0.05(6)   & $J^z_{\langle i, j \rangle}/J_{\langle i, j \rangle}^{xy}$   &  1.37(2) \\

 \end{tabular}
 \end{ruledtabular}
 \end{table}

\appendix
\newpage
\section{Experimental details}
Large single crystals of \cas\ were grown out of excess Sb.  High-purity elements were placed in alumina crucibles with an atomic ratio of Ce 0.045 Ag 0.091 Sb 0.864 and sealed in evacuated amorphous silica tubes. The ampules were heated to 1180~C$^{\circ}$ and then cooled over 100 hours to 670~C$^{\circ}$ after which the excess Sb was decanted with the help of a centrifuge~\cite{canfield2020}. Single crystals with masses as large as $\approx$ 2 grams were grown (see  reference~\cite{myers99} for a representative picture). The sample used in INS experiment consisted of two co-aligned single crystals with total mass, $m = 3.12$~g. The crystals were oriented using the ALF neutron Laue instrument at the ISIS facility. No splitting of the reflections due to twinning or additional reflections arising from secondary grains was observed.

The INS experiments were performed at two time-of-flight spectrometers: Cold Neutron Chopper Spectrometer (CNCS)~\cite{CNCS1,CNCS2}, at the Spallation Neutron Source (SNS) at Oak Ridge National Laboratory and LET spectrometer~\cite{bewley2011let} at the ISIS neutron and muon source.
In the LET experiment~\cite{LET_data} we aligned the samples with the [010]-axis pointed vertically ($(H0L)$ scattering plane) and applied magnetic fields up to 3~T using a vertical cryomagnet. Taking advantage of multirepitition mode we collected the data with three incident neutron energies $E_{\mathrm{i}} = 14.9, 6.5$ and 3.63~meV with energy resolutions of $\Delta{}E = 540, 160$ and 65~$\mu$eV respectively.
The CNCS measurements were performed in two scattering geometries with $(H0L)$ and $(HK0)$ planes of crystal lying in the equatorial plane of the instrument. In both cases magnetic field was applied along the vertical direction using an 8~T cryomagnet. The incident neutron energy was fixed to $E_{\mathrm{i}} = 3.32$~meV ($\Delta{}E = 90~\mu$eV at the elastic line).

Magnetic field dependence of the gap at $\mathbf{Q} = (101)$ was measured at the FLEXX three-axis spectrometer~\cite{le2013gains} at Helmholtz Zentrum Berlin. We have performed measurements in the $(H0L)$ scattering plane, a magnetic field was applied along the $b$ axis using the VM-4 vertical cryomagnet. Measurements were carried out with fixed scattered momentum $k_{\mathrm{f}} = 1.3$~\AA$^{-1}$. A cold beryllium filter was placed between the sample and the analyzer to suppress higher-order contamination of the neutron beam.

Magnetization and the specific heat were measured using MPMS-3 VSM magnetometer and Physical Property Measurement System (PPMS) correspondingly. When the ordered moment of a ferromagetic crystal is oriented perpendicular to the magnetic field, a torque acting on the sample attempts to rotate the moment in the direction of the field, which could lead to a misorientation of the sample in PPMS. We note that the critical field obtained from the specific heat measurements is very close to the one deduced from magnetization measurements in which the sample's orientation was fixed by mounting the sample on a quartz rod. We conclude that a rotation of the sample in specific heat experiments was minimal at fields below 3 T, at higher fields the ordered moment was parallel to the field.

Reduction of the TOF data was done using the $\textsc{Mantid}$~\cite{Mantid} and $\textsc{Horace}$~\cite{Horace} software packages. The calculations of spin dynamics were performed using $\textsc{SpinW}$ package~\cite{toth15}.

\section{Field-induced suppression of the spectral intensity} \label{Sec:Suppression}

Now we would like to discuss the field-induced suppression of the spectral intensity.
To understand the reason for the reduction of the spectral intensity we rewrite the INS cross-section Eq.~\eqref{Eq:INS_cross_section_1} in the following form:
\begin{eqnarray}
\frac{d^2\sigma}{dEd\Omega} \propto \sum_{\alpha, \beta}g_{\alpha}g_{\beta}(\delta_{\alpha,\beta} - \frac{q_{\alpha}q_{\beta}}{q^2})S^{\alpha,\beta},
\label{Eq:INS_cross_section_2}
\end{eqnarray}
where $\alpha, \beta = x,y,z$ and $S^{\alpha\beta}$ is the dynamical structure factor. For the collinear ferromagnet ordered along the $z$-axis, we can reduce Eq.~\eqref{Eq:INS_cross_section_2} to:
\begin{eqnarray}
\frac{d^2\sigma}{dEd\Omega} \propto (g_{xx}^2S^{xx} + g_{yy}^2S^{yy})(1 + \frac{q_{z}^2}{q^2}).
\label{Eq:INS_cross_section_3}
\end{eqnarray}

We consider two cases: (i) at ambient magnetic field, the Ce moments in \cas\ are ordered along the $c$ axis; (ii) when field applied along the $a$-axis and above the $B_{\mathrm{c}}$ the moments are also aligned along [100]. Thus, we can rewrite Eq.~\eqref{Eq:INS_cross_section_3} for these cases, taking into account that $g_{x} \approx 3$, $g_{z} = 1$ for $\Gamma_6$ doublet and $S^{xx} = S^{yy}$:
\begin{eqnarray}
\frac{d^2\sigma}{dEd\Omega} \propto 18S^{xx}(1 + \frac{q_{z}^2}{q^2});\\
\frac{d^2\sigma}{dEd\Omega} \propto (9S^{xx} +S^{zz}) (1 + \frac{q_{y}^2}{q^2}).
\label{cross3}
\end{eqnarray}
Assuming that the $S^{xx} \simeq S^{zz}$, the $g$-factor anisotropy alone accounts for $\sim55$~\% reduction of the spectral intensity, when the spins are reoriented towards the $a$-axis.
Moreover, when the spectra are collected in $(H0L)$ scattering plane at zero field, the polarization factor increase the scattered intensity even more for $\mathbf{q} \parallel q_z$, while when the moments are polarized by a magnetic field, perpendicular to the scattering plane, the polarization factor is 1 for all the in-plane directions. This simple analysis qualitatively describes the origin of the decrease of the spectral intensity.

\section{Signal broadening due to finite $\mathbf{Q}$ integration } \label{Sec:Fat_magnon}

\begin{figure}[b]
\includegraphics[width=.66\linewidth]{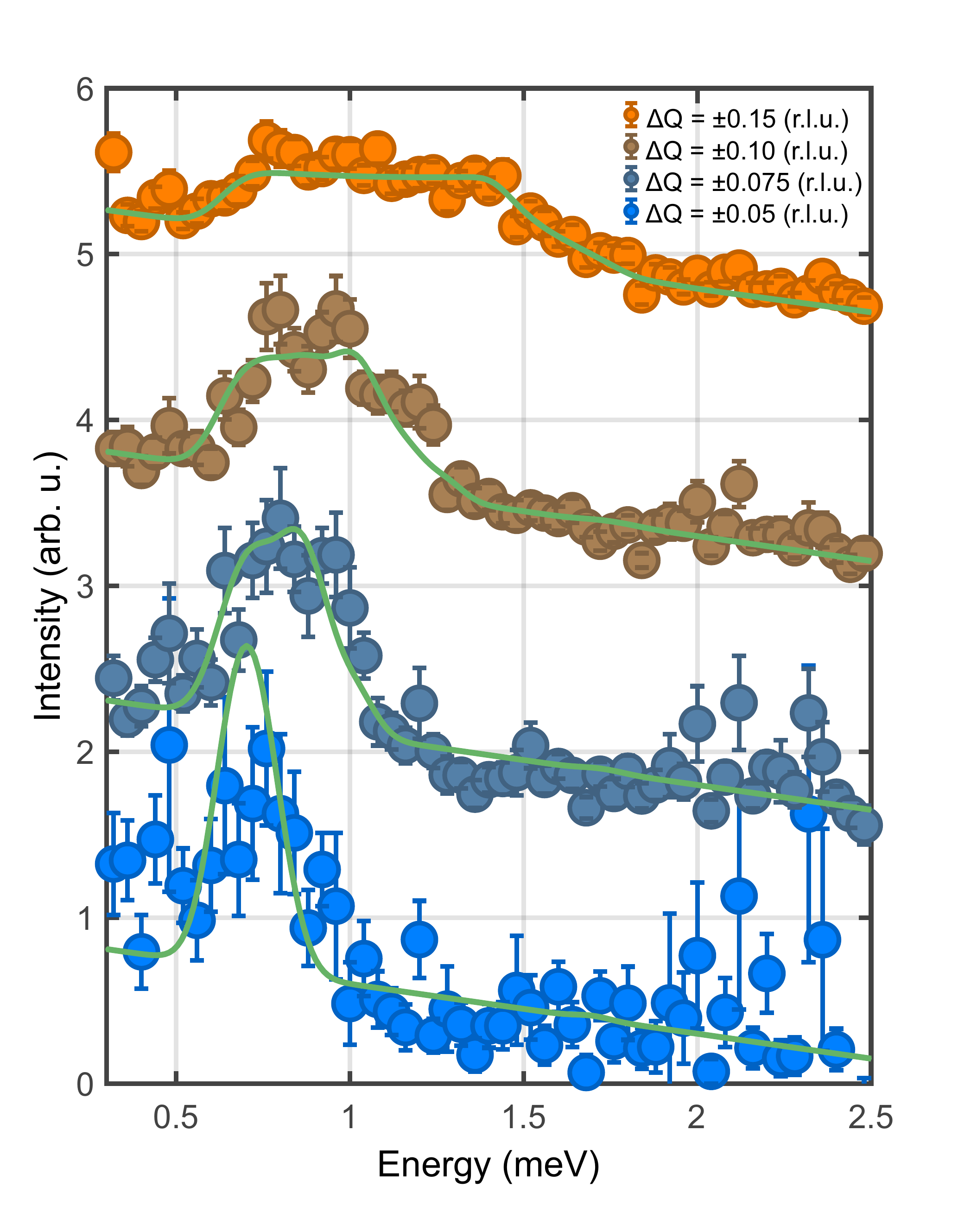}
\caption{~INS spectra of \cas\ measured using CNCS spectrometer at $T = 1.7$~K and $B = 0$~T obtained by integration of the signal around $\Gamma~(101)$-point, $H = 1\pm\Delta{}Q, K~=~0\pm\Delta{}Q, L~=~1\pm\Delta{}Q$ with different $\Delta{}Q$ as indicated in the legend. Green lines show model calculations using SpinW software (on top of an empirical linear background), which take into account $Q$-integration and energy-dependent experimental resolution of CNCS instrument.
The data are shifted vertically for clarity.
}
\label{Fig:Width}
\end{figure}

Figures~\ref{Fig:ZeroFieldMagnons} and \ref{Fig:LETMagnons} show magnon excitations, which have large width when are plotted against $L$-direction of the reciprocal space. Here we performed simple quantitative analysis to present that this broadening is caused by $\mathbf{Q}$-resolution and a finite integration width, when cutting the data from four-dimensional $S(\mathbf{Q},\hbar\omega)$ dataset for the spaghetti plots.

Figure~\ref{Fig:Width} shows INS spectra at $\Gamma$-point $\mathbf{Q} = (101)$ obtained by integration of the zero-field CNCS dataset close to the $\Gamma$ point, $\mathbf{Q} = (101)$ within several different $\mathbf{Q}$-volumes as indicated in the legend. The bright blue dots show signal obtained by integration within the smallest $\mathbf{Q}$-volume and one can see that the data are rather noisy and show a peak at $E \approx 0.6$~meV.
Increase of the integration range apparently enhances the statistics, but also broadens significantly the peak and shifts it towards higher energies, because we integrate in the high-energy magnons  away from the $\Gamma$ point.
To check whether these effects can be captured by our model we calculated the energy cuts taking into account the experimental $\omega$-dependent energy resolution of CNCS as:
\begin{eqnarray*}
     S_{\mathbf{Q} = (101)}(\hbar\omega) = \frac{1}{n_{\scalebox{0.5}{$h$}} n_{\scalebox{0.5}{$k$}} n_{\scalebox{0.5}{$l$}}} \sum \hspace{-0.6cm} \sum_{\scalebox{0.5}{$h, k, l = -\Delta{}Q,...,\Delta{}Q$}}\hspace{-0.6cm}\sum S[h, k, l, \hbar\omega]
\end{eqnarray*}
where $h,k,l$ are reciprocal lattice coordinates, binned on $n_h=n_k=n_l=30$-points 3D-grid between minimal and maximal values $\pm{}\Delta{}Q$.
All calculated curves are multiplied by a constant prefactor to fit the experimental intensity scale plus an empirical linear background are shown by green lines in Fig.~\ref{Fig:Width}. It is clear that three curves calculated for $\Delta{}Q \geq 0.075$ r.l.u. very well reproduce both shape and intensity of the measured excitations. On the other hand, the data for $\Delta{}Q = 0.05$ r.l.u. are somewhat broader and less intense than the calculated spectrum. This is likely to be due to finite intrinsic $\mathbf{Q}$-resolution of the spectrometer, which is of the order of 0.04~\AA$^{-1}$ (0.03 r.l.u. along $(100)$), as estimated from the width of elastic peaks. These number are comparable with the integration width, however we did not account for this broadening in the calculations, because the resolution function of the spectrometer is rather anisotropic and strongly depends on geometry of the sample, spectrometer settings and other minor experimental details.
Thus, we conclude that the large width of the signal in some path of Figs.~\ref{Fig:ZeroFieldMagnons} and \ref{Fig:LETMagnons} is the result of large integration width, which was required to obtain reasonable signal-to-noise ratio.

\section{Comparison of $S = 1/2$ model and full spin Hamiltonian} \label{Sec:mcphase}

In the main text we used $S = 1/2$ model in order to describe low-energy spin dynamics in \cas\ and neglected the excited states. In order to further validate use of the $S = 1/2$ model we performed numerical simulation of INS spectra of the full spin Hamiltonian:
\begin{eqnarray}
    \mathrm{H} = \sum_{\langle i, l, m \rangle} B_l^m O_l^m(\mathbf{J}_i) + \sum_{\langle i, j \rangle} \hat{J}_{ij} \mathbf{J}_i \mathbf{J}_j
    \label{Eq:McPhase_Hamiltonian}
\end{eqnarray}
where $B_l^m$ and $O_l^m$ are Stevens coefficients and operators, $\mathbf{J}$ is the angular momentum operator ($\mathbf{J} = 5/2$ for Ce$^{3+}$ ion), $\hat{J_{ij}}$ are exchange matrix, which couple different pairs of Ce ions.

We used \textsc{McPhase} software~\cite{McPhase1, McPhase2} in order to simulate INS spectrum of \cas\ by means of random phase approximation (RPA). In our simulations we used set of $B_l^m$ parameters deduced in~\cite{araki03} and exchange interactions from Tab.~\ref{tab:Exhange}. It is worth noting that our parameters are given for $S = 1/2$ model and in order to scale them to $\mathbf{J} = 5/2$ we have to multiply $J_{xy}$ components by $1/g_{xy}^2$.

\begin{figure}[b]
\includegraphics[width=.9\linewidth]{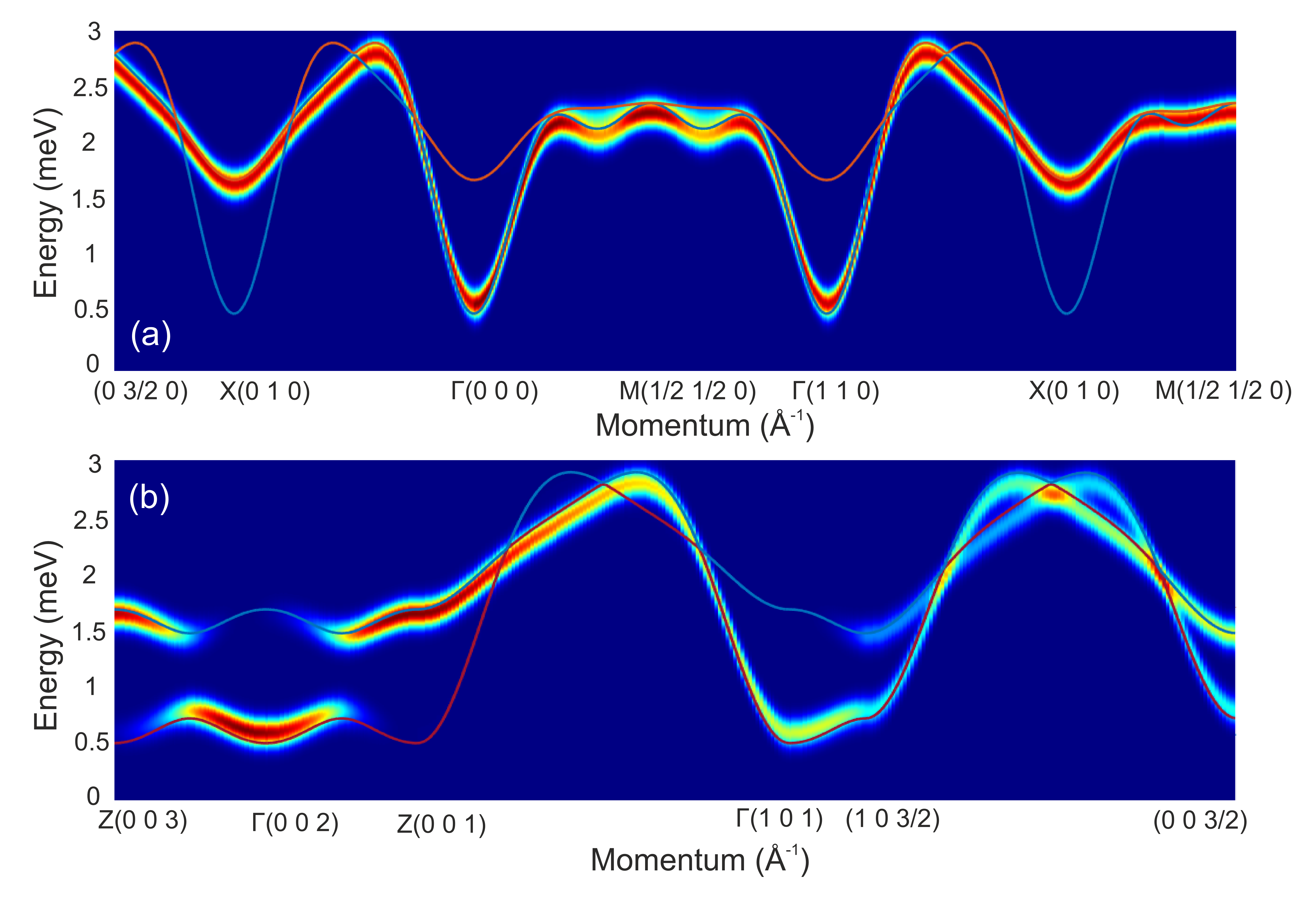}
\caption{~INS spectra calculated by RPA and INS. The colorplots were obtained using \textsc{SpinW} software for the $S = 1/2$ model and are similar to Fig.~\ref{Fig:ZeroFieldMagnons} (b,d). Solid lines show low-energy magnon dispersion calculated for Hamiltonian~\eqref{Eq:McPhase_Hamiltonian} using RPA with \textsc{McPhase} software.
}
\label{Fig:Mcphase_Spinw}
\end{figure}

The low-energy zero-field spectra were calculated using LSWT and RPA for Hamiltonians~\eqref{Eq:SpinHamoltonian} and \eqref{Eq:McPhase_Hamiltonian} respectively and are shown in Fig.~\ref{Fig:Mcphase_Spinw}. Both calculations show very similar dispersion and magnon band width. However, the  dispersion curves calculated by RPA were shifted by 0.15~meV towards higher energy compared to our LSWT results. Therefore for the presented RPA calculations we refined new $J_z/J_{xy} = 1.28$, instead of 1.37 obtained by LSWT to obtain a better agreement.

As a result of this analysis we conclude that both approaches provide almost identical description of the low-energy spin dynamics with the same set of exchange constants justifying use of $S = 1/2$ Hamiltonian in the main text. We further speculate that the minor quantitative disagreement of $J_z/J_{xy}$ values obtained by LSWT and RPA can be either (i) a consequence of difference between the calculation methods, which are known provide different results in some cases; (ii) weak influence of the CEF, which was neglected in LSWT.

\section*{Acknowledgments}
We acknowledge A. S. Sukhanov, O. Stockert, F. Kr\"{u}ger, A. Madsen and C. Geibel for stimulating discussion.
This research used resources at the Spallation Neutron Source, a DOE Office of Science User Facility operated by Oak Ridge National Laboratory.
S.E.N. acknowledges support from the International Max Planck Research School for Chemistry and Physics of Quantum Materials (IMPRS-CPQM) and funding from the European Union’s Horizon 2020 research and innovation program under the Marie Sk{\l}odowska-Curie grant agreement No 884104.
P.C.C. and S.L.B. were supported by the U.S. Department of Energy, Office of Basic Energy Science, Division of Materials Sciences and Engineering. Crystal growth and basic characterization was performed at the Ames Laboratory. Ames Laboratory is operated for the U.S. Department of Energy by Iowa State University under Contract No. DE-AC02-07CH11358.
We gratefully acknowledge the Science and Technology Facilities Council (STFC) for access to neutron beamtime at ISIS.
\newpage
\bibliography{CeAgSb2}

\begin{thebibliography}{48}%
\makeatletter
\providecommand \@ifxundefined [1]{%
 \@ifx{#1\undefined}
}%
\providecommand \@ifnum [1]{%
 \ifnum #1\expandafter \@firstoftwo
 \else \expandafter \@secondoftwo
 \fi
}%
\providecommand \@ifx [1]{%
 \ifx #1\expandafter \@firstoftwo
 \else \expandafter \@secondoftwo
 \fi
}%
\providecommand \natexlab [1]{#1}%
\providecommand \enquote  [1]{``#1''}%
\providecommand \bibnamefont  [1]{#1}%
\providecommand \bibfnamefont [1]{#1}%
\providecommand \citenamefont [1]{#1}%
\providecommand \href@noop [0]{\@secondoftwo}%
\providecommand \href [0]{\begingroup \@sanitize@url \@href}%
\providecommand \@href[1]{\@@startlink{#1}\@@href}%
\providecommand \@@href[1]{\endgroup#1\@@endlink}%
\providecommand \@sanitize@url [0]{\catcode `\\12\catcode `\$12\catcode
  `\&12\catcode `\#12\catcode `\^12\catcode `\_12\catcode `\%12\relax}%
\providecommand \@@startlink[1]{}%
\providecommand \@@endlink[0]{}%
\providecommand \url  [0]{\begingroup\@sanitize@url \@url }%
\providecommand \@url [1]{\endgroup\@href {#1}{\urlprefix }}%
\providecommand \urlprefix  [0]{URL }%
\providecommand \Eprint [0]{\href }%
\providecommand \doibase [0]{https://doi.org/}%
\providecommand \selectlanguage [0]{\@gobble}%
\providecommand \bibinfo  [0]{\@secondoftwo}%
\providecommand \bibfield  [0]{\@secondoftwo}%
\providecommand \translation [1]{[#1]}%
\providecommand \BibitemOpen [0]{}%
\providecommand \bibitemStop [0]{}%
\providecommand \bibitemNoStop [0]{.\EOS\space}%
\providecommand \EOS [0]{\spacefactor3000\relax}%
\providecommand \BibitemShut  [1]{\csname bibitem#1\endcsname}%
\let\auto@bib@innerbib\@empty
%</preamble>
\bibitem [{\citenamefont {Ruderman}\ and\ \citenamefont
  {Kittel}(1954)}]{ruderman1954indirect}%
  \BibitemOpen
  \bibfield  {author} {\bibinfo {author} {\bibfnamefont {M.~A.}\ \bibnamefont
  {Ruderman}}\ and\ \bibinfo {author} {\bibfnamefont {C.}~\bibnamefont
  {Kittel}},\ }\bibfield  {title} {\bibinfo {title} {Indirect exchange coupling
  of nuclear magnetic moments by conduction electrons},\ }\href
  {https://doi.org/10.1103/PhysRev.96.99} {\bibfield  {journal} {\bibinfo
  {journal} {Physical Review}\ }\textbf {\bibinfo {volume} {96}},\ \bibinfo
  {pages} {99} (\bibinfo {year} {1954})}\BibitemShut {NoStop}%
\bibitem [{\citenamefont {Kasuya}(1956)}]{kasuya1956theory}%
  \BibitemOpen
  \bibfield  {author} {\bibinfo {author} {\bibfnamefont {T.}~\bibnamefont
  {Kasuya}},\ }\bibfield  {title} {\bibinfo {title} {{A theory of metallic
  ferro-and antiferromagnetism on Zener's model}},\ }\href
  {https://doi.org/10.1143/PTP.16.45} {\bibfield  {journal} {\bibinfo
  {journal} {Prog. Theor. Phys.}\ }\textbf {\bibinfo {volume} {16}},\ \bibinfo
  {pages} {45} (\bibinfo {year} {1956})}\BibitemShut {NoStop}%
\bibitem [{\citenamefont {Yosida}(1957)}]{yosida1957}%
  \BibitemOpen
  \bibfield  {author} {\bibinfo {author} {\bibfnamefont {K.}~\bibnamefont
  {Yosida}},\ }\bibfield  {title} {\bibinfo {title} {{Magnetic Properties of
  Cu-Mn Alloys}},\ }\href {https://doi.org/10.1103/PhysRev.106.893} {\bibfield
  {journal} {\bibinfo  {journal} {Phys. Rev.}\ }\textbf {\bibinfo {volume}
  {106}},\ \bibinfo {pages} {893} (\bibinfo {year} {1957})}\BibitemShut
  {NoStop}%
\bibitem [{\citenamefont {Mattis}\ and\ \citenamefont
  {Donath}(1962)}]{mattis1962role}%
  \BibitemOpen
  \bibfield  {author} {\bibinfo {author} {\bibfnamefont {D.}~\bibnamefont
  {Mattis}}\ and\ \bibinfo {author} {\bibfnamefont {W.~E.}\ \bibnamefont
  {Donath}},\ }\bibfield  {title} {\bibinfo {title} {Role of fermi surface and
  crystal structure in theory of magnetic metals},\ }\href
  {https://doi.org/10.1103/PhysRev.128.1618} {\bibfield  {journal} {\bibinfo
  {journal} {Phys. Rev.}\ }\textbf {\bibinfo {volume} {128}},\ \bibinfo {pages}
  {1618} (\bibinfo {year} {1962})}\BibitemShut {NoStop}%
\bibitem [{\citenamefont {Hafner}\ \emph {et~al.}(2019)\citenamefont {Hafner},
  \citenamefont {Rai}, \citenamefont {Banda}, \citenamefont {Kliemt},
  \citenamefont {Krellner}, \citenamefont {Sichelschmidt}, \citenamefont
  {Morosan}, \citenamefont {Geibel},\ and\ \citenamefont {Brando}}]{hafner19}%
  \BibitemOpen
  \bibfield  {author} {\bibinfo {author} {\bibfnamefont {D.}~\bibnamefont
  {Hafner}}, \bibinfo {author} {\bibfnamefont {B.~K.}\ \bibnamefont {Rai}},
  \bibinfo {author} {\bibfnamefont {J.}~\bibnamefont {Banda}}, \bibinfo
  {author} {\bibfnamefont {K.}~\bibnamefont {Kliemt}}, \bibinfo {author}
  {\bibfnamefont {C.}~\bibnamefont {Krellner}}, \bibinfo {author}
  {\bibfnamefont {J.}~\bibnamefont {Sichelschmidt}}, \bibinfo {author}
  {\bibfnamefont {E.}~\bibnamefont {Morosan}}, \bibinfo {author} {\bibfnamefont
  {C.}~\bibnamefont {Geibel}},\ and\ \bibinfo {author} {\bibfnamefont
  {M.}~\bibnamefont {Brando}},\ }\bibfield  {title} {\bibinfo {title}
  {{Kondo-lattice ferromagnets and their peculiar order along the magnetically
  hard axis determined by the crystalline electric field}},\ }\href
  {https://doi.org/10.1103/PhysRevB.99.201109} {\bibfield  {journal} {\bibinfo
  {journal} {Phys. Rev. B}\ }\textbf {\bibinfo {volume} {99}},\ \bibinfo
  {pages} {201109} (\bibinfo {year} {2019})}\BibitemShut {NoStop}%
\bibitem [{\citenamefont {Ahamed}\ \emph {et~al.}(2018)\citenamefont {Ahamed},
  \citenamefont {Moessner},\ and\ \citenamefont {Erten}}]{ahamed2018rare}%
  \BibitemOpen
  \bibfield  {author} {\bibinfo {author} {\bibfnamefont {S.}~\bibnamefont
  {Ahamed}}, \bibinfo {author} {\bibfnamefont {R.}~\bibnamefont {Moessner}},\
  and\ \bibinfo {author} {\bibfnamefont {O.}~\bibnamefont {Erten}},\ }\bibfield
   {title} {\bibinfo {title} {{Why rare-earth ferromagnets are so rare:
  Insights from the $p$-wave Kondo model}},\ }\href
  {https://doi.org/10.1103/PhysRevB.98.054420} {\bibfield  {journal} {\bibinfo
  {journal} {Phys. Rev. B}\ }\textbf {\bibinfo {volume} {98}},\ \bibinfo
  {pages} {054420} (\bibinfo {year} {2018})}\BibitemShut {NoStop}%
\bibitem [{\citenamefont {Brando}\ \emph {et~al.}(2016)\citenamefont {Brando},
  \citenamefont {Belitz}, \citenamefont {Grosche},\ and\ \citenamefont
  {Kirkpatrick}}]{brando2016metallic}%
  \BibitemOpen
  \bibfield  {author} {\bibinfo {author} {\bibfnamefont {M.}~\bibnamefont
  {Brando}}, \bibinfo {author} {\bibfnamefont {D.}~\bibnamefont {Belitz}},
  \bibinfo {author} {\bibfnamefont {F.~M.}\ \bibnamefont {Grosche}},\ and\
  \bibinfo {author} {\bibfnamefont {T.~R.}\ \bibnamefont {Kirkpatrick}},\
  }\bibfield  {title} {\bibinfo {title} {Metallic quantum ferromagnets},\
  }\href {https://doi.org/10.1103/RevModPhys.88.025006} {\bibfield  {journal}
  {\bibinfo  {journal} {Rev. Mod. Phys.}\ }\textbf {\bibinfo {volume} {88}},\
  \bibinfo {pages} {025006} (\bibinfo {year} {2016})}\BibitemShut {NoStop}%
\bibitem [{\citenamefont {Houshiar}\ \emph {et~al.}(1995)\citenamefont
  {Houshiar}, \citenamefont {Adroja},\ and\ \citenamefont
  {Rainford}}]{HOUSHIAR19951231}%
  \BibitemOpen
  \bibfield  {author} {\bibinfo {author} {\bibfnamefont {M.}~\bibnamefont
  {Houshiar}}, \bibinfo {author} {\bibfnamefont {D.}~\bibnamefont {Adroja}},\
  and\ \bibinfo {author} {\bibfnamefont {B.}~\bibnamefont {Rainford}},\
  }\bibfield  {title} {\bibinfo {title} {{Kondo lattice behaviour in CeTSb$_2$
  compounds (T=Ni, Cu and Ag)}},\ }\href
  {https://doi.org/https://doi.org/10.1016/0304-8853(94)01418-3} {\bibfield
  {journal} {\bibinfo  {journal} {J. Magn. Magn. Mater.}\ }\textbf {\bibinfo
  {volume} {140-144}},\ \bibinfo {pages} {1231} (\bibinfo {year}
  {1995})}\BibitemShut {NoStop}%
\bibitem [{\citenamefont {Myers}\ \emph {et~al.}(1999)\citenamefont {Myers},
  \citenamefont {Bud'ko}, \citenamefont {Fisher}, \citenamefont {Islam},
  \citenamefont {Kleinke}, \citenamefont {Lacerda},\ and\ \citenamefont
  {Canfield}}]{myers99}%
  \BibitemOpen
  \bibfield  {author} {\bibinfo {author} {\bibfnamefont {K.}~\bibnamefont
  {Myers}}, \bibinfo {author} {\bibfnamefont {S.}~\bibnamefont {Bud'ko}},
  \bibinfo {author} {\bibfnamefont {I.}~\bibnamefont {Fisher}}, \bibinfo
  {author} {\bibfnamefont {Z.}~\bibnamefont {Islam}}, \bibinfo {author}
  {\bibfnamefont {H.}~\bibnamefont {Kleinke}}, \bibinfo {author} {\bibfnamefont
  {A.}~\bibnamefont {Lacerda}},\ and\ \bibinfo {author} {\bibfnamefont
  {P.}~\bibnamefont {Canfield}},\ }\bibfield  {title} {\bibinfo {title}
  {{Systematic study of anisotropic transport and magnetic properties of
  RAgSb$_2$ (R=Y, La–Nd, Sm, Gd–Tm)}},\ }\href
  {https://doi.org/https://doi.org/10.1016/S0304-8853(99)00472-2} {\bibfield
  {journal} {\bibinfo  {journal} {J. Magn. Magn. Mater.}\ }\textbf {\bibinfo
  {volume} {205}},\ \bibinfo {pages} {27 } (\bibinfo {year}
  {1999})}\BibitemShut {NoStop}%
\bibitem [{\citenamefont {Jobiliong}\ \emph {et~al.}(2005)\citenamefont
  {Jobiliong}, \citenamefont {Brooks}, \citenamefont {Choi}, \citenamefont
  {Lee},\ and\ \citenamefont {Fisk}}]{jobiliong2005}%
  \BibitemOpen
  \bibfield  {author} {\bibinfo {author} {\bibfnamefont {E.}~\bibnamefont
  {Jobiliong}}, \bibinfo {author} {\bibfnamefont {J.~S.}\ \bibnamefont
  {Brooks}}, \bibinfo {author} {\bibfnamefont {E.~S.}\ \bibnamefont {Choi}},
  \bibinfo {author} {\bibfnamefont {H.}~\bibnamefont {Lee}},\ and\ \bibinfo
  {author} {\bibfnamefont {Z.}~\bibnamefont {Fisk}},\ }\bibfield  {title}
  {\bibinfo {title} {{Magnetization and electrical-transport investigation of
  the dense Kondo system $\mathrm{Ce}\mathrm{Ag}{\mathrm{Sb}}_{2}$}},\ }\href
  {https://doi.org/10.1103/PhysRevB.72.104428} {\bibfield  {journal} {\bibinfo
  {journal} {Phys. Rev. B}\ }\textbf {\bibinfo {volume} {72}},\ \bibinfo
  {pages} {104428} (\bibinfo {year} {2005})}\BibitemShut {NoStop}%
\bibitem [{\citenamefont {Takeuchi}\ \emph {et~al.}(2003)\citenamefont
  {Takeuchi}, \citenamefont {Thamizhavel}, \citenamefont {Okubo}, \citenamefont
  {Yamada}, \citenamefont {Nakamura}, \citenamefont {Yamamoto}, \citenamefont
  {Inada}, \citenamefont {Sugiyama}, \citenamefont {Galatanu}, \citenamefont
  {Yamamoto}, \citenamefont {Kindo}, \citenamefont {Ebihara},\ and\
  \citenamefont {\ifmmode~\bar{O}\else \={O}\fi{}nuki}}]{takeuchi2003}%
  \BibitemOpen
  \bibfield  {author} {\bibinfo {author} {\bibfnamefont {T.}~\bibnamefont
  {Takeuchi}}, \bibinfo {author} {\bibfnamefont {A.}~\bibnamefont
  {Thamizhavel}}, \bibinfo {author} {\bibfnamefont {T.}~\bibnamefont {Okubo}},
  \bibinfo {author} {\bibfnamefont {M.}~\bibnamefont {Yamada}}, \bibinfo
  {author} {\bibfnamefont {N.}~\bibnamefont {Nakamura}}, \bibinfo {author}
  {\bibfnamefont {T.}~\bibnamefont {Yamamoto}}, \bibinfo {author}
  {\bibfnamefont {Y.}~\bibnamefont {Inada}}, \bibinfo {author} {\bibfnamefont
  {K.}~\bibnamefont {Sugiyama}}, \bibinfo {author} {\bibfnamefont
  {A.}~\bibnamefont {Galatanu}}, \bibinfo {author} {\bibfnamefont
  {E.}~\bibnamefont {Yamamoto}}, \bibinfo {author} {\bibfnamefont
  {K.}~\bibnamefont {Kindo}}, \bibinfo {author} {\bibfnamefont
  {T.}~\bibnamefont {Ebihara}},\ and\ \bibinfo {author} {\bibfnamefont
  {Y.}~\bibnamefont {\ifmmode~\bar{O}\else \={O}\fi{}nuki}},\ }\bibfield
  {title} {\bibinfo {title} {{Anisotropic, thermal, and magnetic properties of
  ${\mathrm{CeAgSb}}_{2}:$ Explanation via a crystalline electric field
  scheme}},\ }\href {https://doi.org/10.1103/PhysRevB.67.064403} {\bibfield
  {journal} {\bibinfo  {journal} {Phys. Rev. B}\ }\textbf {\bibinfo {volume}
  {67}},\ \bibinfo {pages} {064403} (\bibinfo {year} {2003})}\BibitemShut
  {NoStop}%
\bibitem [{\citenamefont {Araki}\ \emph {et~al.}(2003)\citenamefont {Araki},
  \citenamefont {Metoki}, \citenamefont {Galatanu}, \citenamefont {Yamamoto},
  \citenamefont {Thamizhavel},\ and\ \citenamefont {\ifmmode~\bar{O}\else
  \={O}\fi{}nuki}}]{araki03}%
  \BibitemOpen
  \bibfield  {author} {\bibinfo {author} {\bibfnamefont {S.}~\bibnamefont
  {Araki}}, \bibinfo {author} {\bibfnamefont {N.}~\bibnamefont {Metoki}},
  \bibinfo {author} {\bibfnamefont {A.}~\bibnamefont {Galatanu}}, \bibinfo
  {author} {\bibfnamefont {E.}~\bibnamefont {Yamamoto}}, \bibinfo {author}
  {\bibfnamefont {A.}~\bibnamefont {Thamizhavel}},\ and\ \bibinfo {author}
  {\bibfnamefont {Y.}~\bibnamefont {\ifmmode~\bar{O}\else \={O}\fi{}nuki}},\
  }\bibfield  {title} {\bibinfo {title} {{Crystal structure, magnetic ordering,
  and magnetic excitation in the $4f$-localized ferromagnet
  ${\mathrm{CeAgSb}}_{2}$}},\ }\href
  {https://doi.org/10.1103/PhysRevB.68.024408} {\bibfield  {journal} {\bibinfo
  {journal} {Phys. Rev. B}\ }\textbf {\bibinfo {volume} {68}},\ \bibinfo
  {pages} {024408} (\bibinfo {year} {2003})}\BibitemShut {NoStop}%
\bibitem [{\citenamefont {Saitoh}\ \emph {et~al.}(2016)\citenamefont {Saitoh},
  \citenamefont {Fujiwara}, \citenamefont {Yamaguchi}, \citenamefont
  {Nakatani}, \citenamefont {Mori}, \citenamefont {Fuchimoto}, \citenamefont
  {Kiss}, \citenamefont {Yasui}, \citenamefont {Miyawaki}, \citenamefont
  {Imada}, \citenamefont {Yamagami}, \citenamefont {Ebihara},\ and\
  \citenamefont {Sekiyama}}]{saitoh2016}%
  \BibitemOpen
  \bibfield  {author} {\bibinfo {author} {\bibfnamefont {Y.}~\bibnamefont
  {Saitoh}}, \bibinfo {author} {\bibfnamefont {H.}~\bibnamefont {Fujiwara}},
  \bibinfo {author} {\bibfnamefont {T.}~\bibnamefont {Yamaguchi}}, \bibinfo
  {author} {\bibfnamefont {Y.}~\bibnamefont {Nakatani}}, \bibinfo {author}
  {\bibfnamefont {T.}~\bibnamefont {Mori}}, \bibinfo {author} {\bibfnamefont
  {H.}~\bibnamefont {Fuchimoto}}, \bibinfo {author} {\bibfnamefont
  {T.}~\bibnamefont {Kiss}}, \bibinfo {author} {\bibfnamefont {A.}~\bibnamefont
  {Yasui}}, \bibinfo {author} {\bibfnamefont {J.}~\bibnamefont {Miyawaki}},
  \bibinfo {author} {\bibfnamefont {S.}~\bibnamefont {Imada}}, \bibinfo
  {author} {\bibfnamefont {H.}~\bibnamefont {Yamagami}}, \bibinfo {author}
  {\bibfnamefont {T.}~\bibnamefont {Ebihara}},\ and\ \bibinfo {author}
  {\bibfnamefont {A.}~\bibnamefont {Sekiyama}},\ }\bibfield  {title} {\bibinfo
  {title} {{Electronic Structures of Ferromagnetic CeAgSb$_2$: Soft X-ray
  Absorption, Magnetic Circular Dichroism, and Angle-Resolved Photoemission
  Spectroscopies}},\ }\href {https://doi.org/10.7566/JPSJ.85.114713} {\bibfield
   {journal} {\bibinfo  {journal} {J. Phys. Soc. Japan}\ }\textbf {\bibinfo
  {volume} {85}},\ \bibinfo {pages} {114713} (\bibinfo {year}
  {2016})}\BibitemShut {NoStop}%
\bibitem [{\citenamefont {Kr\"uger}\ \emph {et~al.}(2014)\citenamefont
  {Kr\"uger}, \citenamefont {Pedder},\ and\ \citenamefont {Green}}]{krueger14}%
  \BibitemOpen
  \bibfield  {author} {\bibinfo {author} {\bibfnamefont {F.}~\bibnamefont
  {Kr\"uger}}, \bibinfo {author} {\bibfnamefont {C.~J.}\ \bibnamefont
  {Pedder}},\ and\ \bibinfo {author} {\bibfnamefont {A.~G.}\ \bibnamefont
  {Green}},\ }\bibfield  {title} {\bibinfo {title} {{Fluctuation-Driven
  Magnetic Hard-Axis Ordering in Metallic Ferromagnets}},\ }\href
  {https://doi.org/10.1103/PhysRevLett.113.147001} {\bibfield  {journal}
  {\bibinfo  {journal} {Phys. Rev. Lett.}\ }\textbf {\bibinfo {volume} {113}},\
  \bibinfo {pages} {147001} (\bibinfo {year} {2014})}\BibitemShut {NoStop}%
\bibitem [{\citenamefont {Kwasigroch}\ \emph {et~al.}(2020)\citenamefont
  {Kwasigroch}, \citenamefont {Hu}, \citenamefont {Krüger},\ and\
  \citenamefont {Green}}]{kwasigroch2020magnetic}%
  \BibitemOpen
  \bibfield  {author} {\bibinfo {author} {\bibfnamefont {M.~P.}\ \bibnamefont
  {Kwasigroch}}, \bibinfo {author} {\bibfnamefont {H.}~\bibnamefont {Hu}},
  \bibinfo {author} {\bibfnamefont {F.}~\bibnamefont {Krüger}},\ and\ \bibinfo
  {author} {\bibfnamefont {A.~G.}\ \bibnamefont {Green}},\ }\href@noop {}
  {\bibinfo {title} {{Magnetic hard-direction ordering in anisotropic Kondo
  systems}}} (\bibinfo {year} {2020}),\ \Eprint
  {https://arxiv.org/abs/2005.09533} {arXiv:2005.09533 [cond-mat.str-el]}
  \BibitemShut {NoStop}%
\bibitem [{\citenamefont {Logg}\ \emph {et~al.}(2013)\citenamefont {Logg},
  \citenamefont {Feng}, \citenamefont {Ebihara}, \citenamefont {Zou},
  \citenamefont {Friedemann}, \citenamefont {Alireza}, \citenamefont {Goh},\
  and\ \citenamefont {Grosche}}]{logg2013}%
  \BibitemOpen
  \bibfield  {author} {\bibinfo {author} {\bibfnamefont {P.}~\bibnamefont
  {Logg}}, \bibinfo {author} {\bibfnamefont {Z.}~\bibnamefont {Feng}}, \bibinfo
  {author} {\bibfnamefont {T.}~\bibnamefont {Ebihara}}, \bibinfo {author}
  {\bibfnamefont {Y.}~\bibnamefont {Zou}}, \bibinfo {author} {\bibfnamefont
  {S.}~\bibnamefont {Friedemann}}, \bibinfo {author} {\bibfnamefont
  {P.}~\bibnamefont {Alireza}}, \bibinfo {author} {\bibfnamefont
  {S.}~\bibnamefont {Goh}},\ and\ \bibinfo {author} {\bibfnamefont {F.~M.}\
  \bibnamefont {Grosche}},\ }\bibfield  {title} {\bibinfo {title} {{Pressure
  and field tuning in the Kondo lattice ferromagnet CeAgSb$_2$}},\ }\href
  {https://doi.org/10.1002/pssb.201200924} {\bibfield  {journal} {\bibinfo
  {journal} {Phys. Status Solidi (b)}\ }\textbf {\bibinfo {volume} {250}},\
  \bibinfo {pages} {515} (\bibinfo {year} {2013})}\BibitemShut {NoStop}%
\bibitem [{\citenamefont {Kawasaki}\ \emph {et~al.}(2018)\citenamefont
  {Kawasaki}, \citenamefont {Ogata}, \citenamefont {Kawai}, \citenamefont
  {Fukuyama}, \citenamefont {Yamaguchi},\ and\ \citenamefont
  {Sumiyama}}]{kawasaki2018}%
  \BibitemOpen
  \bibfield  {author} {\bibinfo {author} {\bibfnamefont {I.}~\bibnamefont
  {Kawasaki}}, \bibinfo {author} {\bibfnamefont {S.}~\bibnamefont {Ogata}},
  \bibinfo {author} {\bibfnamefont {S.}~\bibnamefont {Kawai}}, \bibinfo
  {author} {\bibfnamefont {Y.}~\bibnamefont {Fukuyama}}, \bibinfo {author}
  {\bibfnamefont {A.}~\bibnamefont {Yamaguchi}},\ and\ \bibinfo {author}
  {\bibfnamefont {A.}~\bibnamefont {Sumiyama}},\ }\bibfield  {title} {\bibinfo
  {title} {{Magnetic Properties of 4$f$ Localized Ferromagnet CeAgSb$_2$ under
  Transverse Magnetic Fields}},\ }\href
  {https://doi.org/10.7566/JPSJ.87.014703} {\bibfield  {journal} {\bibinfo
  {journal} {J. Phys. Soc. Japan}\ }\textbf {\bibinfo {volume} {87}},\ \bibinfo
  {pages} {014703} (\bibinfo {year} {2018})}\BibitemShut {NoStop}%
\bibitem [{\citenamefont {Portnichenko}\ \emph {et~al.}(2016)\citenamefont
  {Portnichenko}, \citenamefont {Romh\'{a}nyi}, \citenamefont {Onykiienko},
  \citenamefont {Henschel}, \citenamefont {Schmidt}, \citenamefont {Cameron},
  \citenamefont {Surmach}, \citenamefont {Lim}, \citenamefont {Park},
  \citenamefont {Schneidewind}, \citenamefont {Abernathy}, \citenamefont
  {Rosner}, \citenamefont {van~den Brink},\ and\ \citenamefont
  {Inosov}}]{portnichenko2016magnon}%
  \BibitemOpen
  \bibfield  {author} {\bibinfo {author} {\bibfnamefont {P.~Y.}\ \bibnamefont
  {Portnichenko}}, \bibinfo {author} {\bibfnamefont {J.}~\bibnamefont
  {Romh\'{a}nyi}}, \bibinfo {author} {\bibfnamefont {Y.~A.}\ \bibnamefont
  {Onykiienko}}, \bibinfo {author} {\bibfnamefont {A.}~\bibnamefont
  {Henschel}}, \bibinfo {author} {\bibfnamefont {M.}~\bibnamefont {Schmidt}},
  \bibinfo {author} {\bibfnamefont {A.~S.}\ \bibnamefont {Cameron}}, \bibinfo
  {author} {\bibfnamefont {M.~A.}\ \bibnamefont {Surmach}}, \bibinfo {author}
  {\bibfnamefont {J.~A.}\ \bibnamefont {Lim}}, \bibinfo {author} {\bibfnamefont
  {J.~T.}\ \bibnamefont {Park}}, \bibinfo {author} {\bibfnamefont
  {A.}~\bibnamefont {Schneidewind}}, \bibinfo {author} {\bibfnamefont {D.~L.}\
  \bibnamefont {Abernathy}}, \bibinfo {author} {\bibfnamefont {H.}~\bibnamefont
  {Rosner}}, \bibinfo {author} {\bibfnamefont {J.}~\bibnamefont {van~den
  Brink}},\ and\ \bibinfo {author} {\bibfnamefont {D.~S.}\ \bibnamefont
  {Inosov}},\ }\bibfield  {title} {\bibinfo {title} {{Magnon spectrum of the
  helimagnetic insulator Cu$_2$OSeO$_3$}},\ }\href
  {https://doi.org/10.1038/ncomms10725} {\bibfield  {journal} {\bibinfo
  {journal} {Nat. Commun.}\ }\textbf {\bibinfo {volume} {7}},\ \bibinfo {pages}
  {1} (\bibinfo {year} {2016})}\BibitemShut {NoStop}%
\bibitem [{\citenamefont {Tymoshenko}\ \emph {et~al.}(2017)\citenamefont
  {Tymoshenko}, \citenamefont {Onykiienko}, \citenamefont {M{\"u}ller},
  \citenamefont {Thomale}, \citenamefont {Rachel}, \citenamefont {Cameron},
  \citenamefont {Portnichenko}, \citenamefont {Efremov}, \citenamefont
  {Tsurkan}, \citenamefont {Abernathy}, \citenamefont {Ollivier}, \citenamefont
  {Schneidewind}, \citenamefont {Piovano}, \citenamefont {Felea}, \citenamefont
  {Loidl},\ and\ \citenamefont {Inosov}}]{tymoshenko2017pseudo}%
  \BibitemOpen
  \bibfield  {author} {\bibinfo {author} {\bibfnamefont {Y.~V.}\ \bibnamefont
  {Tymoshenko}}, \bibinfo {author} {\bibfnamefont {Y.~A.}\ \bibnamefont
  {Onykiienko}}, \bibinfo {author} {\bibfnamefont {T.}~\bibnamefont
  {M{\"u}ller}}, \bibinfo {author} {\bibfnamefont {R.}~\bibnamefont {Thomale}},
  \bibinfo {author} {\bibfnamefont {S.}~\bibnamefont {Rachel}}, \bibinfo
  {author} {\bibfnamefont {A.~S.}\ \bibnamefont {Cameron}}, \bibinfo {author}
  {\bibfnamefont {P.~Y.}\ \bibnamefont {Portnichenko}}, \bibinfo {author}
  {\bibfnamefont {D.~V.}\ \bibnamefont {Efremov}}, \bibinfo {author}
  {\bibfnamefont {V.}~\bibnamefont {Tsurkan}}, \bibinfo {author} {\bibfnamefont
  {D.~L.}\ \bibnamefont {Abernathy}}, \bibinfo {author} {\bibfnamefont
  {J.}~\bibnamefont {Ollivier}}, \bibinfo {author} {\bibfnamefont
  {A.}~\bibnamefont {Schneidewind}}, \bibinfo {author} {\bibfnamefont
  {A.}~\bibnamefont {Piovano}}, \bibinfo {author} {\bibfnamefont
  {V.}~\bibnamefont {Felea}}, \bibinfo {author} {\bibfnamefont
  {A.}~\bibnamefont {Loidl}},\ and\ \bibinfo {author} {\bibfnamefont
  {D.}~\bibnamefont {Inosov}},\ }\bibfield  {title} {\bibinfo {title}
  {{Pseudo-Goldstone Magnons in the Frustrated $S= 3/2$ Heisenberg Helimagnet
  ZnCr$_2$Se$_4$ with a Pyrochlore Magnetic Sublattice}},\ }\href
  {https://doi.org/10.1103/PhysRevX.7.041049} {\bibfield  {journal} {\bibinfo
  {journal} {Phys. Rev. X}\ }\textbf {\bibinfo {volume} {7}},\ \bibinfo {pages}
  {041049} (\bibinfo {year} {2017})}\BibitemShut {NoStop}%
\bibitem [{\citenamefont {Wu}\ \emph {et~al.}(2016)\citenamefont {Wu},
  \citenamefont {Gannon}, \citenamefont {Zaliznyak}, \citenamefont {Tsvelik},
  \citenamefont {Brockmann}, \citenamefont {Caux}, \citenamefont {Kim},
  \citenamefont {Qiu}, \citenamefont {Copley}, \citenamefont {Ehlers},
  \citenamefont {Podlesnyak},\ and\ \citenamefont {Aronson}}]{Wu2016}%
  \BibitemOpen
  \bibfield  {author} {\bibinfo {author} {\bibfnamefont {L.~S.}\ \bibnamefont
  {Wu}}, \bibinfo {author} {\bibfnamefont {W.~J.}\ \bibnamefont {Gannon}},
  \bibinfo {author} {\bibfnamefont {I.~A.}\ \bibnamefont {Zaliznyak}}, \bibinfo
  {author} {\bibfnamefont {A.~M.}\ \bibnamefont {Tsvelik}}, \bibinfo {author}
  {\bibfnamefont {M.}~\bibnamefont {Brockmann}}, \bibinfo {author}
  {\bibfnamefont {J.-S.}\ \bibnamefont {Caux}}, \bibinfo {author}
  {\bibfnamefont {M.~S.}\ \bibnamefont {Kim}}, \bibinfo {author} {\bibfnamefont
  {Y.}~\bibnamefont {Qiu}}, \bibinfo {author} {\bibfnamefont {J.~R.~D.}\
  \bibnamefont {Copley}}, \bibinfo {author} {\bibfnamefont {G.}~\bibnamefont
  {Ehlers}}, \bibinfo {author} {\bibfnamefont {A.}~\bibnamefont {Podlesnyak}},\
  and\ \bibinfo {author} {\bibfnamefont {M.~C.}\ \bibnamefont {Aronson}},\
  }\bibfield  {title} {\bibinfo {title} {{Orbital-exchange and fractional
  quantum number excitations in an f-electron metal, Yb$_2$Pt$_2$Pb}},\ }\href
  {https://doi.org/10.1126/science.aaf0981} {\bibfield  {journal} {\bibinfo
  {journal} {Science}\ }\textbf {\bibinfo {volume} {352}},\ \bibinfo {pages}
  {1206} (\bibinfo {year} {2016})}\BibitemShut {NoStop}%
\bibitem [{\citenamefont {Paddison}\ \emph {et~al.}(2017)\citenamefont
  {Paddison}, \citenamefont {Daum}, \citenamefont {Dun}, \citenamefont
  {Ehlers}, \citenamefont {Liu}, \citenamefont {Stone}, \citenamefont {Zhou},\
  and\ \citenamefont {Mourigal}}]{paddison2017continuous}%
  \BibitemOpen
  \bibfield  {author} {\bibinfo {author} {\bibfnamefont {J.~A.~M.}\
  \bibnamefont {Paddison}}, \bibinfo {author} {\bibfnamefont {M.}~\bibnamefont
  {Daum}}, \bibinfo {author} {\bibfnamefont {Z.}~\bibnamefont {Dun}}, \bibinfo
  {author} {\bibfnamefont {G.}~\bibnamefont {Ehlers}}, \bibinfo {author}
  {\bibfnamefont {Y.}~\bibnamefont {Liu}}, \bibinfo {author} {\bibfnamefont
  {M.~B.}\ \bibnamefont {Stone}}, \bibinfo {author} {\bibfnamefont
  {H.}~\bibnamefont {Zhou}},\ and\ \bibinfo {author} {\bibfnamefont
  {M.}~\bibnamefont {Mourigal}},\ }\bibfield  {title} {\bibinfo {title}
  {{Continuous excitations of the triangular-lattice quantum spin liquid
  YbMgGaO$_4$}},\ }\href {https://doi.org/10.1038/nphys3971} {\bibfield
  {journal} {\bibinfo  {journal} {Nat. Phys.}\ }\textbf {\bibinfo {volume}
  {13}},\ \bibinfo {pages} {117} (\bibinfo {year} {2017})}\BibitemShut
  {NoStop}%
\bibitem [{\citenamefont {Nikitin}\ \emph {et~al.}(2018)\citenamefont
  {Nikitin}, \citenamefont {Wu}, \citenamefont {Sefat}, \citenamefont
  {Shaykhutdinov}, \citenamefont {Lu}, \citenamefont {Meng}, \citenamefont
  {Pomjakushina}, \citenamefont {Conder}, \citenamefont {Ehlers}, \citenamefont
  {Lumsden}, \citenamefont {Kolesnikov}, \citenamefont {Barilo}, \citenamefont
  {Guretskii}, \citenamefont {Inosov},\ and\ \citenamefont
  {Podlesnyak}}]{Nikitin2018}%
  \BibitemOpen
  \bibfield  {author} {\bibinfo {author} {\bibfnamefont {S.~E.}\ \bibnamefont
  {Nikitin}}, \bibinfo {author} {\bibfnamefont {L.~S.}\ \bibnamefont {Wu}},
  \bibinfo {author} {\bibfnamefont {A.~S.}\ \bibnamefont {Sefat}}, \bibinfo
  {author} {\bibfnamefont {K.~A.}\ \bibnamefont {Shaykhutdinov}}, \bibinfo
  {author} {\bibfnamefont {Z.}~\bibnamefont {Lu}}, \bibinfo {author}
  {\bibfnamefont {S.}~\bibnamefont {Meng}}, \bibinfo {author} {\bibfnamefont
  {E.~V.}\ \bibnamefont {Pomjakushina}}, \bibinfo {author} {\bibfnamefont
  {K.}~\bibnamefont {Conder}}, \bibinfo {author} {\bibfnamefont
  {G.}~\bibnamefont {Ehlers}}, \bibinfo {author} {\bibfnamefont {M.~D.}\
  \bibnamefont {Lumsden}}, \bibinfo {author} {\bibfnamefont {A.~I.}\
  \bibnamefont {Kolesnikov}}, \bibinfo {author} {\bibfnamefont
  {S.}~\bibnamefont {Barilo}}, \bibinfo {author} {\bibfnamefont {S.~A.}\
  \bibnamefont {Guretskii}}, \bibinfo {author} {\bibfnamefont {D.~S.}\
  \bibnamefont {Inosov}},\ and\ \bibinfo {author} {\bibfnamefont
  {A.}~\bibnamefont {Podlesnyak}},\ }\bibfield  {title} {\bibinfo {title}
  {{Decoupled spin dynamics in the rare-earth orthoferrite
  ${\mathrm{YbFeO}}_{3}$: Evolution of magnetic excitations through the
  spin-reorientation transition}},\ }\href
  {https://doi.org/10.1103/PhysRevB.98.064424} {\bibfield  {journal} {\bibinfo
  {journal} {Phys. Rev. B}\ }\textbf {\bibinfo {volume} {98}},\ \bibinfo
  {pages} {064424} (\bibinfo {year} {2018})}\BibitemShut {NoStop}%
\bibitem [{\citenamefont {Wu}\ \emph {et~al.}(2019)\citenamefont {Wu},
  \citenamefont {Nikitin}, \citenamefont {Wang}, \citenamefont {Zhu},
  \citenamefont {Batista}, \citenamefont {Tsvelik}, \citenamefont {Samarakoon},
  \citenamefont {Tennant}, \citenamefont {Brando}, \citenamefont {Vasylechko},
  \citenamefont {Frontzek}, \citenamefont {Savici}, \citenamefont {Sala},
  \citenamefont {Ehlers}, \citenamefont {Christianson}, \citenamefont
  {Lumsden},\ and\ \citenamefont {Podlesnyak}}]{Wu2019}%
  \BibitemOpen
  \bibfield  {author} {\bibinfo {author} {\bibfnamefont {L.~S.}\ \bibnamefont
  {Wu}}, \bibinfo {author} {\bibfnamefont {S.~E.}\ \bibnamefont {Nikitin}},
  \bibinfo {author} {\bibfnamefont {Z.}~\bibnamefont {Wang}}, \bibinfo {author}
  {\bibfnamefont {W.}~\bibnamefont {Zhu}}, \bibinfo {author} {\bibfnamefont
  {C.~D.}\ \bibnamefont {Batista}}, \bibinfo {author} {\bibfnamefont {A.~M.}\
  \bibnamefont {Tsvelik}}, \bibinfo {author} {\bibfnamefont {A.~M.}\
  \bibnamefont {Samarakoon}}, \bibinfo {author} {\bibfnamefont {D.~A.}\
  \bibnamefont {Tennant}}, \bibinfo {author} {\bibfnamefont {M.}~\bibnamefont
  {Brando}}, \bibinfo {author} {\bibfnamefont {L.}~\bibnamefont {Vasylechko}},
  \bibinfo {author} {\bibfnamefont {M.}~\bibnamefont {Frontzek}}, \bibinfo
  {author} {\bibfnamefont {A.~T.}\ \bibnamefont {Savici}}, \bibinfo {author}
  {\bibfnamefont {G.}~\bibnamefont {Sala}}, \bibinfo {author} {\bibfnamefont
  {G.}~\bibnamefont {Ehlers}}, \bibinfo {author} {\bibfnamefont {A.~D.}\
  \bibnamefont {Christianson}}, \bibinfo {author} {\bibfnamefont {M.~D.}\
  \bibnamefont {Lumsden}},\ and\ \bibinfo {author} {\bibfnamefont
  {A.}~\bibnamefont {Podlesnyak}},\ }\bibfield  {title} {\bibinfo {title}
  {{Tomonaga-Luttinger liquid behavior and spinon confinement in YbAlO$_3$}},\
  }\href {https://doi.org/10.1038/s41467-019-08485-7} {\bibfield  {journal}
  {\bibinfo  {journal} {Nat. Commun.}\ }\textbf {\bibinfo {volume} {10}},\
  \bibinfo {pages} {698} (\bibinfo {year} {2019})}\BibitemShut {NoStop}%
\bibitem [{\citenamefont {Gao}\ \emph {et~al.}(2019)\citenamefont {Gao},
  \citenamefont {Chen}, \citenamefont {Tam}, \citenamefont {Huang},
  \citenamefont {Sasmal}, \citenamefont {Adroja}, \citenamefont {Ye},
  \citenamefont {Cao}, \citenamefont {Sala}, \citenamefont {Stone},
  \citenamefont {Baines}, \citenamefont {Verezhak}, \citenamefont {Hu},
  \citenamefont {Chung}, \citenamefont {Xu}, \citenamefont {Cheong},
  \citenamefont {Nallaiyan}, \citenamefont {Spagna}, \citenamefont {Maple},
  \citenamefont {Nevidomskyy}, \citenamefont {Morosan}, \citenamefont {Chen},\
  and\ \citenamefont {Dai}}]{gao2019experimental}%
  \BibitemOpen
  \bibfield  {author} {\bibinfo {author} {\bibfnamefont {B.}~\bibnamefont
  {Gao}}, \bibinfo {author} {\bibfnamefont {T.}~\bibnamefont {Chen}}, \bibinfo
  {author} {\bibfnamefont {D.~W.}\ \bibnamefont {Tam}}, \bibinfo {author}
  {\bibfnamefont {C.-L.}\ \bibnamefont {Huang}}, \bibinfo {author}
  {\bibfnamefont {K.}~\bibnamefont {Sasmal}}, \bibinfo {author} {\bibfnamefont
  {D.~T.}\ \bibnamefont {Adroja}}, \bibinfo {author} {\bibfnamefont
  {F.}~\bibnamefont {Ye}}, \bibinfo {author} {\bibfnamefont {H.}~\bibnamefont
  {Cao}}, \bibinfo {author} {\bibfnamefont {G.}~\bibnamefont {Sala}}, \bibinfo
  {author} {\bibfnamefont {M.~B.}\ \bibnamefont {Stone}}, \bibinfo {author}
  {\bibfnamefont {C.}~\bibnamefont {Baines}}, \bibinfo {author} {\bibfnamefont
  {J.~A.~T.}\ \bibnamefont {Verezhak}}, \bibinfo {author} {\bibfnamefont
  {H.}~\bibnamefont {Hu}}, \bibinfo {author} {\bibfnamefont {J.-H.}\
  \bibnamefont {Chung}}, \bibinfo {author} {\bibfnamefont {X.}~\bibnamefont
  {Xu}}, \bibinfo {author} {\bibfnamefont {S.-W.}\ \bibnamefont {Cheong}},
  \bibinfo {author} {\bibfnamefont {M.}~\bibnamefont {Nallaiyan}}, \bibinfo
  {author} {\bibfnamefont {S.}~\bibnamefont {Spagna}}, \bibinfo {author}
  {\bibfnamefont {M.~B.}\ \bibnamefont {Maple}}, \bibinfo {author}
  {\bibfnamefont {A.~H.}\ \bibnamefont {Nevidomskyy}}, \bibinfo {author}
  {\bibfnamefont {E.}~\bibnamefont {Morosan}}, \bibinfo {author} {\bibfnamefont
  {G.}~\bibnamefont {Chen}},\ and\ \bibinfo {author} {\bibfnamefont
  {P.}~\bibnamefont {Dai}},\ }\bibfield  {title} {\bibinfo {title}
  {{Experimental signatures of a three-dimensional quantum spin liquid in
  effective spin-1/2 Ce$_2$Zr$_2$O$_7$ pyrochlore}},\ }\href
  {https://doi.org/10.1038/s41567-019-0577-6} {\bibfield  {journal} {\bibinfo
  {journal} {Nature Phys.}\ }\textbf {\bibinfo {volume} {15}},\ \bibinfo
  {pages} {1052} (\bibinfo {year} {2019})}\BibitemShut {NoStop}%
\bibitem [{\citenamefont {Toth}\ and\ \citenamefont {Lake}(2015)}]{toth15}%
  \BibitemOpen
  \bibfield  {author} {\bibinfo {author} {\bibfnamefont {S.}~\bibnamefont
  {Toth}}\ and\ \bibinfo {author} {\bibfnamefont {B.}~\bibnamefont {Lake}},\
  }\bibfield  {title} {\bibinfo {title} {{Linear spin wave theory for
  single-{Q} incommensurate magnetic structures}},\ }\href
  {https://doi.org/10.1088/0953-8984/27/16/166002} {\bibfield  {journal}
  {\bibinfo  {journal} {J. Phys.: Condens. Matter}\ }\textbf {\bibinfo {volume}
  {27}},\ \bibinfo {pages} {166002} (\bibinfo {year} {2015})}\BibitemShut
  {NoStop}%
\bibitem [{\citenamefont {Squires}(1978)}]{Squires_book_1978}%
  \BibitemOpen
  \bibfield  {author} {\bibinfo {author} {\bibfnamefont {G.~L.}\ \bibnamefont
  {Squires}},\ }\href@noop {} {\emph {\bibinfo {title} {Introduction to the
  Theory of Thermal Neutron Scattering}}},\ \bibinfo {edition} {2012th}\ ed.\
  (\bibinfo  {publisher} {Cambridge University Press},\ \bibinfo {address}
  {England},\ \bibinfo {year} {1978})\BibitemShut {NoStop}%
\bibitem [{\citenamefont {Lovesey}(1984)}]{lovesey1984theory}%
  \BibitemOpen
  \bibfield  {author} {\bibinfo {author} {\bibfnamefont {S.~W.}\ \bibnamefont
  {Lovesey}},\ }\bibinfo {title} {Theory of neutron scattering from condensed
  matter}\ (\bibinfo  {publisher} {Oxford Science Publishers},\ \bibinfo {year}
  {1984})\BibitemShut {NoStop}%
\bibitem [{\citenamefont {Zaliznyak}\ and\ \citenamefont
  {Lee}(2005)}]{ZaliznyakLee_MNSChapter}%
  \BibitemOpen
  \bibfield  {author} {\bibinfo {author} {\bibfnamefont {I.~A.}\ \bibnamefont
  {Zaliznyak}}\ and\ \bibinfo {author} {\bibfnamefont {S.-H.}\ \bibnamefont
  {Lee}},\ }\bibfield  {title} {\bibinfo {title} {Magnetic neutron
  scattering},\ }in\ \href {https://doi.org/10.1007/0-387-23395-4_1} {\emph
  {\bibinfo {booktitle} {Modern Techniques for Characterizing Magnetic
  Materials}}},\ \bibinfo {editor} {edited by\ \bibinfo {editor} {\bibfnamefont
  {Y.}~\bibnamefont {Zhu}}}\ (\bibinfo  {publisher} {Springer US},\ \bibinfo
  {year} {2005})\ pp.\ \bibinfo {pages} {3--64}\BibitemShut {NoStop}%
\bibitem [{Note1()}]{Note1}%
  \BibitemOpen
  \bibinfo {note} {We note that these slices were obtained by $|\protect
  \mathbf {Q}|$-averaging the signal within the $(H0L)$ scattering plane rather
  than the full reciprocal space, and therefore somewhat differ from a real
  powder spectra.}\BibitemShut {Stop}%
\bibitem [{Note2()}]{Note2}%
  \BibitemOpen
  \bibinfo {note} {We discuss a possible origin of the suppression of the
  spectral intensity in Sec.~\ref {Sec:Suppression}}\BibitemShut {NoStop}%
\bibitem [{\citenamefont {Pankrats}\ \emph {et~al.}(2004)\citenamefont
  {Pankrats}, \citenamefont {Tugarinov},\ and\ \citenamefont
  {Sablina}}]{pankrats2004magnetic}%
  \BibitemOpen
  \bibfield  {author} {\bibinfo {author} {\bibfnamefont {A.~I.}\ \bibnamefont
  {Pankrats}}, \bibinfo {author} {\bibfnamefont {V.~I.}\ \bibnamefont
  {Tugarinov}},\ and\ \bibinfo {author} {\bibfnamefont {K.~A.}\ \bibnamefont
  {Sablina}},\ }\bibfield  {title} {\bibinfo {title} {{Magnetic resonance in
  new copper oxide Cu$_5$Bi$_2$B$_4$O$_{14}$ with triclinic symmetry}},\ }\href
  {https://doi.org/10.1016/j.jmmm.2004.01.083} {\bibfield  {journal} {\bibinfo
  {journal} {J. Magn. Magn. Mater.}\ }\textbf {\bibinfo {volume} {279}},\
  \bibinfo {pages} {231} (\bibinfo {year} {2004})}\BibitemShut {NoStop}%
\bibitem [{\citenamefont {Selter}\ \emph {et~al.}(2020)\citenamefont {Selter},
  \citenamefont {Bastien}, \citenamefont {Wolter}, \citenamefont {Aswartham},\
  and\ \citenamefont {B{\"u}chner}}]{selter2020magnetic}%
  \BibitemOpen
  \bibfield  {author} {\bibinfo {author} {\bibfnamefont {S.}~\bibnamefont
  {Selter}}, \bibinfo {author} {\bibfnamefont {G.}~\bibnamefont {Bastien}},
  \bibinfo {author} {\bibfnamefont {A.~U.~B.}\ \bibnamefont {Wolter}}, \bibinfo
  {author} {\bibfnamefont {S.}~\bibnamefont {Aswartham}},\ and\ \bibinfo
  {author} {\bibfnamefont {B.}~\bibnamefont {B{\"u}chner}},\ }\bibfield
  {title} {\bibinfo {title} {{Magnetic anisotropy and low-field magnetic phase
  diagram of the quasi-two-dimensional ferromagnet Cr$_2$Ge$_2$Te$_6$}},\
  }\href {https://doi.org/10.1103/PhysRevB.101.014440} {\bibfield  {journal}
  {\bibinfo  {journal} {Phys. Rev. B}\ }\textbf {\bibinfo {volume} {101}},\
  \bibinfo {pages} {014440} (\bibinfo {year} {2020})}\BibitemShut {NoStop}%
\bibitem [{\citenamefont {Pankrats}\ \emph {et~al.}(2016)\citenamefont
  {Pankrats}, \citenamefont {Sablina}, \citenamefont {Eremin}, \citenamefont
  {Balaev}, \citenamefont {Kolkov}, \citenamefont {Tugarinov},\ and\
  \citenamefont {Bovina}}]{pankrats2016ferromagnetism}%
  \BibitemOpen
  \bibfield  {author} {\bibinfo {author} {\bibfnamefont {A.}~\bibnamefont
  {Pankrats}}, \bibinfo {author} {\bibfnamefont {K.}~\bibnamefont {Sablina}},
  \bibinfo {author} {\bibfnamefont {M.}~\bibnamefont {Eremin}}, \bibinfo
  {author} {\bibfnamefont {A.}~\bibnamefont {Balaev}}, \bibinfo {author}
  {\bibfnamefont {M.}~\bibnamefont {Kolkov}}, \bibinfo {author} {\bibfnamefont
  {V.}~\bibnamefont {Tugarinov}},\ and\ \bibinfo {author} {\bibfnamefont
  {A.}~\bibnamefont {Bovina}},\ }\bibfield  {title} {\bibinfo {title}
  {{Ferromagnetism and strong magnetic anisotropy of the PbMnBO$_4$ orthoborate
  single crystals}},\ }\href {https://doi.org/10.1016/j.jmmm.2016.04.042}
  {\bibfield  {journal} {\bibinfo  {journal} {J. Magn. Magn. Mater.}\ }\textbf
  {\bibinfo {volume} {414}},\ \bibinfo {pages} {82} (\bibinfo {year}
  {2016})}\BibitemShut {NoStop}%
\bibitem [{\citenamefont {Adroja}\ \emph {et~al.}(2002)\citenamefont {Adroja},
  \citenamefont {Riedi}, \citenamefont {Armitage},\ and\ \citenamefont
  {Fort}}]{adroja2002thermal}%
  \BibitemOpen
  \bibfield  {author} {\bibinfo {author} {\bibfnamefont {D.~T.}\ \bibnamefont
  {Adroja}}, \bibinfo {author} {\bibfnamefont {P.~C.}\ \bibnamefont {Riedi}},
  \bibinfo {author} {\bibfnamefont {J.~G.~M.}\ \bibnamefont {Armitage}},\ and\
  \bibinfo {author} {\bibfnamefont {D.}~\bibnamefont {Fort}},\ }\bibfield
  {title} {\bibinfo {title} {{Thermal Expansion and Magnetostriction Studies of
  a Kondo Lattice Compound: CeAgSb$_2$}},\ }\href
  {https://arxiv.org/abs/cond-mat/0206505} {\bibfield  {journal} {\bibinfo
  {journal} {arXiv preprint cond-mat/0206505}\ } (\bibinfo {year}
  {2002})}\BibitemShut {NoStop}%
\bibitem [{\citenamefont {Endoh}\ \emph {et~al.}(1984)\citenamefont {Endoh},
  \citenamefont {Ajiro}, \citenamefont {Shiba},\ and\ \citenamefont
  {Yoshizawa}}]{endoh1984resonant}%
  \BibitemOpen
  \bibfield  {author} {\bibinfo {author} {\bibfnamefont {Y.}~\bibnamefont
  {Endoh}}, \bibinfo {author} {\bibfnamefont {Y.}~\bibnamefont {Ajiro}},
  \bibinfo {author} {\bibfnamefont {H.}~\bibnamefont {Shiba}},\ and\ \bibinfo
  {author} {\bibfnamefont {H.}~\bibnamefont {Yoshizawa}},\ }\bibfield  {title}
  {\bibinfo {title} {Resonant coupling between one-and two-magnon excitations
  in tetramethylamine manganese trichloride (tmmc)},\ }\href
  {https://doi.org/10.1103/PhysRevB.30.4074} {\bibfield  {journal} {\bibinfo
  {journal} {Phys. Rev. B}\ }\textbf {\bibinfo {volume} {30}},\ \bibinfo
  {pages} {4074} (\bibinfo {year} {1984})}\BibitemShut {NoStop}%
\bibitem [{\citenamefont {Kuroe}\ \emph {et~al.}(2011)\citenamefont {Kuroe},
  \citenamefont {Hamasaki}, \citenamefont {Sekine}, \citenamefont {Hase},
  \citenamefont {Oka}, \citenamefont {Ito}, \citenamefont {Eisaki},
  \citenamefont {Kaneko}, \citenamefont {Metoki}, \citenamefont {Matsuda},\
  and\ \citenamefont {Kakurai}}]{kuroe2011hybridization}%
  \BibitemOpen
  \bibfield  {author} {\bibinfo {author} {\bibfnamefont {H.}~\bibnamefont
  {Kuroe}}, \bibinfo {author} {\bibfnamefont {T.}~\bibnamefont {Hamasaki}},
  \bibinfo {author} {\bibfnamefont {T.}~\bibnamefont {Sekine}}, \bibinfo
  {author} {\bibfnamefont {M.}~\bibnamefont {Hase}}, \bibinfo {author}
  {\bibfnamefont {K.}~\bibnamefont {Oka}}, \bibinfo {author} {\bibfnamefont
  {T.}~\bibnamefont {Ito}}, \bibinfo {author} {\bibfnamefont {H.}~\bibnamefont
  {Eisaki}}, \bibinfo {author} {\bibfnamefont {K.}~\bibnamefont {Kaneko}},
  \bibinfo {author} {\bibfnamefont {N.}~\bibnamefont {Metoki}}, \bibinfo
  {author} {\bibfnamefont {M.}~\bibnamefont {Matsuda}},\ and\ \bibinfo {author}
  {\bibfnamefont {K.}~\bibnamefont {Kakurai}},\ }\bibfield  {title} {\bibinfo
  {title} {{Hybridization of magnetic excitations between quasi-one-dimensional
  spin chains and spin dimers in Cu$_3$Mo$_2$O$_9$ observed using inelastic
  neutron scattering}},\ }\href {https://doi.org/10.1103/PhysRevB.83.184423}
  {\bibfield  {journal} {\bibinfo  {journal} {Phys. Rev. B}\ }\textbf {\bibinfo
  {volume} {83}},\ \bibinfo {pages} {184423} (\bibinfo {year}
  {2011})}\BibitemShut {NoStop}%
\bibitem [{\citenamefont {Hayashida}\ \emph {et~al.}(2015)\citenamefont
  {Hayashida}, \citenamefont {Soda}, \citenamefont {Itoh}, \citenamefont
  {Yokoo}, \citenamefont {Ohgushi}, \citenamefont {Kawana}, \citenamefont
  {R{\o}nnow},\ and\ \citenamefont {Masuda}}]{hayashida2015magnetic}%
  \BibitemOpen
  \bibfield  {author} {\bibinfo {author} {\bibfnamefont {S.}~\bibnamefont
  {Hayashida}}, \bibinfo {author} {\bibfnamefont {M.}~\bibnamefont {Soda}},
  \bibinfo {author} {\bibfnamefont {S.}~\bibnamefont {Itoh}}, \bibinfo {author}
  {\bibfnamefont {T.}~\bibnamefont {Yokoo}}, \bibinfo {author} {\bibfnamefont
  {K.}~\bibnamefont {Ohgushi}}, \bibinfo {author} {\bibfnamefont
  {D.}~\bibnamefont {Kawana}}, \bibinfo {author} {\bibfnamefont {H.~M.}\
  \bibnamefont {R{\o}nnow}},\ and\ \bibinfo {author} {\bibfnamefont
  {T.}~\bibnamefont {Masuda}},\ }\bibfield  {title} {\bibinfo {title}
  {{Magnetic model in multiferroic NdFe$_3$(BO$_3$)$_4$ investigated by
  inelastic neutron scattering}},\ }\href
  {https://doi.org/10.1103/PhysRevB.92.054402} {\bibfield  {journal} {\bibinfo
  {journal} {Phys. Rev. B}\ }\textbf {\bibinfo {volume} {92}},\ \bibinfo
  {pages} {054402} (\bibinfo {year} {2015})}\BibitemShut {NoStop}%
\bibitem [{\citenamefont {Marcus}\ \emph {et~al.}(2018)\citenamefont {Marcus},
  \citenamefont {Kim}, \citenamefont {Tutmaher}, \citenamefont
  {Rodriguez-Rivera}, \citenamefont {Birk}, \citenamefont {Niedermeyer},
  \citenamefont {Lee}, \citenamefont {Fisk},\ and\ \citenamefont
  {Broholm}}]{marcus2018}%
  \BibitemOpen
  \bibfield  {author} {\bibinfo {author} {\bibfnamefont {G.~G.}\ \bibnamefont
  {Marcus}}, \bibinfo {author} {\bibfnamefont {D.-J.}\ \bibnamefont {Kim}},
  \bibinfo {author} {\bibfnamefont {J.~A.}\ \bibnamefont {Tutmaher}}, \bibinfo
  {author} {\bibfnamefont {J.~A.}\ \bibnamefont {Rodriguez-Rivera}}, \bibinfo
  {author} {\bibfnamefont {J.~O.}\ \bibnamefont {Birk}}, \bibinfo {author}
  {\bibfnamefont {C.}~\bibnamefont {Niedermeyer}}, \bibinfo {author}
  {\bibfnamefont {H.}~\bibnamefont {Lee}}, \bibinfo {author} {\bibfnamefont
  {Z.}~\bibnamefont {Fisk}},\ and\ \bibinfo {author} {\bibfnamefont {C.~L.}\
  \bibnamefont {Broholm}},\ }\bibfield  {title} {\bibinfo {title} {{Multi-$q$
  Mesoscale Magnetism in ${\mathrm{CeAuSb}}_{2}$}},\ }\href
  {https://doi.org/10.1103/PhysRevLett.120.097201} {\bibfield  {journal}
  {\bibinfo  {journal} {Phys. Rev. Lett.}\ }\textbf {\bibinfo {volume} {120}},\
  \bibinfo {pages} {097201} (\bibinfo {year} {2018})}\BibitemShut {NoStop}%
\bibitem [{\citenamefont {Canfield}(2020)}]{canfield2020}%
  \BibitemOpen
  \bibfield  {author} {\bibinfo {author} {\bibfnamefont {P.~C.}\ \bibnamefont
  {Canfield}},\ }\bibfield  {title} {\bibinfo {title} {{New materials
  physics}},\ }\href {https://doi.org/10.1088/1361-6633/ab514b} {\bibfield
  {journal} {\bibinfo  {journal} {Rep. Prog. Phys.}\ }\textbf {\bibinfo
  {volume} {83}},\ \bibinfo {pages} {016501} (\bibinfo {year}
  {2020})}\BibitemShut {NoStop}%
\bibitem [{\citenamefont {Ehlers}\ \emph {et~al.}(2011)\citenamefont {Ehlers},
  \citenamefont {Podlesnyak}, \citenamefont {Niedziela}, \citenamefont
  {Iverson},\ and\ \citenamefont {Sokol}}]{CNCS1}%
  \BibitemOpen
  \bibfield  {author} {\bibinfo {author} {\bibfnamefont {G.}~\bibnamefont
  {Ehlers}}, \bibinfo {author} {\bibfnamefont {A.}~\bibnamefont {Podlesnyak}},
  \bibinfo {author} {\bibfnamefont {J.~L.}\ \bibnamefont {Niedziela}}, \bibinfo
  {author} {\bibfnamefont {E.~B.}\ \bibnamefont {Iverson}},\ and\ \bibinfo
  {author} {\bibfnamefont {P.~E.}\ \bibnamefont {Sokol}},\ }\bibfield  {title}
  {\bibinfo {title} {The new cold neutron chopper spectrometer at the
  spallation neutron source: design and performance},\ }\href
  {https://doi.org/10.1063/1.3626935} {\bibfield  {journal} {\bibinfo
  {journal} {Rev. Sci. Instrum.}\ }\textbf {\bibinfo {volume} {82}},\ \bibinfo
  {pages} {085108} (\bibinfo {year} {2011})}\BibitemShut {NoStop}%
\bibitem [{\citenamefont {Ehlers}\ \emph {et~al.}(2016)\citenamefont {Ehlers},
  \citenamefont {Podlesnyak},\ and\ \citenamefont {Kolesnikov}}]{CNCS2}%
  \BibitemOpen
  \bibfield  {author} {\bibinfo {author} {\bibfnamefont {G.}~\bibnamefont
  {Ehlers}}, \bibinfo {author} {\bibfnamefont {A.}~\bibnamefont {Podlesnyak}},\
  and\ \bibinfo {author} {\bibfnamefont {A.~I.}\ \bibnamefont {Kolesnikov}},\
  }\bibfield  {title} {\bibinfo {title} {{The cold neutron chopper spectrometer
  at the Spallation Neutron Source - A review of the first 8 years of
  operation}},\ }\href {https://doi.org/10.1063/1.4962024} {\bibfield
  {journal} {\bibinfo  {journal} {Rev. Sci. Instrum.}\ }\textbf {\bibinfo
  {volume} {87}},\ \bibinfo {pages} {093902} (\bibinfo {year}
  {2016})}\BibitemShut {NoStop}%
\bibitem [{\citenamefont {Bewley}\ \emph {et~al.}(2011)\citenamefont {Bewley},
  \citenamefont {Taylor},\ and\ \citenamefont {Bennington}}]{bewley2011let}%
  \BibitemOpen
  \bibfield  {author} {\bibinfo {author} {\bibfnamefont {R.~I.}\ \bibnamefont
  {Bewley}}, \bibinfo {author} {\bibfnamefont {J.~W.}\ \bibnamefont {Taylor}},\
  and\ \bibinfo {author} {\bibfnamefont {S.~M.}\ \bibnamefont {Bennington}},\
  }\bibfield  {title} {\bibinfo {title} {{LET, a cold neutron multi-disk
  chopper spectrometer at ISIS}},\ }\href
  {https://doi.org/10.1016/j.nima.2011.01.173} {\bibfield  {journal} {\bibinfo
  {journal} {Nucl. Instrum. Methods Phys. Res.}\ }\textbf {\bibinfo {volume}
  {637}},\ \bibinfo {pages} {128} (\bibinfo {year} {2011})}\BibitemShut
  {NoStop}%
\bibitem [{\citenamefont {D.~Sokolov}\ and\ \citenamefont
  {Canfield}()}]{LET_data}%
  \BibitemOpen
  \bibfield  {author} {\bibinfo {author} {\bibfnamefont {D.~V.}\ \bibnamefont
  {D.~Sokolov}, \bibfnamefont {S.~Bud'ko}}\ and\ \bibinfo {author}
  {\bibfnamefont {P.}~\bibnamefont {Canfield}},\ }\bibfield  {title} {\bibinfo
  {title} {{Hard-Axis Ordering in Ferromagnetic CeAgSb$_2$}},\ }\bibfield
  {journal} {\bibinfo  {journal} {STFC ISIS Neutron and Muon Source}\ }\href
  {https://doi.org/10.5286/ISIS.E.RB1520015}
  {10.5286/ISIS.E.RB1520015}\BibitemShut {NoStop}%
\bibitem [{\citenamefont {Le}\ \emph {et~al.}(2013)\citenamefont {Le},
  \citenamefont {Quintero-Castro}, \citenamefont {Toft-Petersen}, \citenamefont
  {Groitl}, \citenamefont {Skoulatos}, \citenamefont {Rule},\ and\
  \citenamefont {Habicht}}]{le2013gains}%
  \BibitemOpen
  \bibfield  {author} {\bibinfo {author} {\bibfnamefont {M.~D.}\ \bibnamefont
  {Le}}, \bibinfo {author} {\bibfnamefont {D.~L.}\ \bibnamefont
  {Quintero-Castro}}, \bibinfo {author} {\bibfnamefont {R.}~\bibnamefont
  {Toft-Petersen}}, \bibinfo {author} {\bibfnamefont {F.}~\bibnamefont
  {Groitl}}, \bibinfo {author} {\bibfnamefont {M.}~\bibnamefont {Skoulatos}},
  \bibinfo {author} {\bibfnamefont {K.}~\bibnamefont {Rule}},\ and\ \bibinfo
  {author} {\bibfnamefont {K.}~\bibnamefont {Habicht}},\ }\bibfield  {title}
  {\bibinfo {title} {{Gains from the upgrade of the cold neutron triple-axis
  spectrometer FLEXX at the BER-II reactor}},\ }\href
  {https://doi.org/10.1016/j.nima.2013.07.007} {\bibfield  {journal} {\bibinfo
  {journal} {Nucl. Instrum. Methods Phys. Res. Sect. A}\ }\textbf {\bibinfo
  {volume} {729}},\ \bibinfo {pages} {220} (\bibinfo {year}
  {2013})}\BibitemShut {NoStop}%
\bibitem [{\citenamefont {Arnold}\ \emph {et~al.}(2014)\citenamefont {Arnold},
  \citenamefont {Bilheux}, \citenamefont {Borreguero}, \citenamefont {Buts},
  \citenamefont {Campbell}, \citenamefont {Chapon}, \citenamefont {Doucet},
  \citenamefont {Draper}, \citenamefont {Leal}, \citenamefont {Gigg},
  \citenamefont {Lynch}, \citenamefont {Markvardsen}, \citenamefont
  {Mikkelson}, \citenamefont {Mikkelson}, \citenamefont {Miller}, \citenamefont
  {Palmen}, \citenamefont {Parker}, \citenamefont {Passos}, \citenamefont
  {Perring}, \citenamefont {Peterson}, \citenamefont {Ren}, \citenamefont
  {Reuter}, \citenamefont {Savici}, \citenamefont {Taylor}, \citenamefont
  {Taylor}, \citenamefont {Tolchenov}, \citenamefont {Zhou},\ and\
  \citenamefont {Zikovsky}}]{Mantid}%
  \BibitemOpen
  \bibfield  {author} {\bibinfo {author} {\bibfnamefont {O.}~\bibnamefont
  {Arnold}}, \bibinfo {author} {\bibfnamefont {J.~C.}\ \bibnamefont {Bilheux}},
  \bibinfo {author} {\bibfnamefont {J.~M.}\ \bibnamefont {Borreguero}},
  \bibinfo {author} {\bibfnamefont {A.}~\bibnamefont {Buts}}, \bibinfo {author}
  {\bibfnamefont {S.~I.}\ \bibnamefont {Campbell}}, \bibinfo {author}
  {\bibfnamefont {L.}~\bibnamefont {Chapon}}, \bibinfo {author} {\bibfnamefont
  {M.}~\bibnamefont {Doucet}}, \bibinfo {author} {\bibfnamefont
  {N.}~\bibnamefont {Draper}}, \bibinfo {author} {\bibfnamefont {R.~F.}\
  \bibnamefont {Leal}}, \bibinfo {author} {\bibfnamefont {M.~A.}\ \bibnamefont
  {Gigg}}, \bibinfo {author} {\bibfnamefont {V.~E.}\ \bibnamefont {Lynch}},
  \bibinfo {author} {\bibfnamefont {A.}~\bibnamefont {Markvardsen}}, \bibinfo
  {author} {\bibfnamefont {D.~J.}\ \bibnamefont {Mikkelson}}, \bibinfo {author}
  {\bibfnamefont {R.~L.}\ \bibnamefont {Mikkelson}}, \bibinfo {author}
  {\bibfnamefont {R.}~\bibnamefont {Miller}}, \bibinfo {author} {\bibfnamefont
  {K.}~\bibnamefont {Palmen}}, \bibinfo {author} {\bibfnamefont
  {P.}~\bibnamefont {Parker}}, \bibinfo {author} {\bibfnamefont
  {G.}~\bibnamefont {Passos}}, \bibinfo {author} {\bibfnamefont {T.~G.}\
  \bibnamefont {Perring}}, \bibinfo {author} {\bibfnamefont {P.~F.}\
  \bibnamefont {Peterson}}, \bibinfo {author} {\bibfnamefont {S.}~\bibnamefont
  {Ren}}, \bibinfo {author} {\bibfnamefont {M.~A.}\ \bibnamefont {Reuter}},
  \bibinfo {author} {\bibfnamefont {A.~T.}\ \bibnamefont {Savici}}, \bibinfo
  {author} {\bibfnamefont {J.~W.}\ \bibnamefont {Taylor}}, \bibinfo {author}
  {\bibfnamefont {R.~J.}\ \bibnamefont {Taylor}}, \bibinfo {author}
  {\bibfnamefont {R.}~\bibnamefont {Tolchenov}}, \bibinfo {author}
  {\bibfnamefont {W.}~\bibnamefont {Zhou}},\ and\ \bibinfo {author}
  {\bibfnamefont {J.}~\bibnamefont {Zikovsky}},\ }\bibfield  {title} {\bibinfo
  {title} {{Mantid -- Data analysis and visualization package for neutron
  scattering and $\mu$SR experiments}},\ }\href
  {https://doi.org/10.1016/j.nima.2014.07.029} {\bibfield  {journal} {\bibinfo
  {journal} {Nucl. Instrum. Methods Phys. Res. Sect. A}\ }\textbf {\bibinfo
  {volume} {764}},\ \bibinfo {pages} {156} (\bibinfo {year}
  {2014})}\BibitemShut {NoStop}%
\bibitem [{\citenamefont {Ewings}\ \emph {et~al.}(2016)\citenamefont {Ewings},
  \citenamefont {Buts}, \citenamefont {Le}, \citenamefont {van Duijn},
  \citenamefont {Bustinduy},\ and\ \citenamefont {Perring}}]{Horace}%
  \BibitemOpen
  \bibfield  {author} {\bibinfo {author} {\bibfnamefont {R.~A.}\ \bibnamefont
  {Ewings}}, \bibinfo {author} {\bibfnamefont {A.}~\bibnamefont {Buts}},
  \bibinfo {author} {\bibfnamefont {M.~D.}\ \bibnamefont {Le}}, \bibinfo
  {author} {\bibfnamefont {J.}~\bibnamefont {van Duijn}}, \bibinfo {author}
  {\bibfnamefont {I.}~\bibnamefont {Bustinduy}},\ and\ \bibinfo {author}
  {\bibfnamefont {T.~G.}\ \bibnamefont {Perring}},\ }\bibfield  {title}
  {\bibinfo {title} {{HORACE: software for the analysis of data from single
  crystal spectroscopy experiments at time-of-flight neutron instruments}},\
  }\href {https://doi.org/10.1016/j.nima.2016.07.036} {\bibfield  {journal}
  {\bibinfo  {journal} {Nucl. Instrum. Methods Phys. Res. Sect. A}\ }\textbf
  {\bibinfo {volume} {834}},\ \bibinfo {pages} {3132} (\bibinfo {year}
  {2016})}\BibitemShut {NoStop}%
\bibitem [{McP()}]{McPhase1}%
  \BibitemOpen
  \href@noop {} {}\bibinfo {howpublished} {http://www.mcphase.de; M. Rotter, J.
  Magn. Magn. Mater. \textbf{272-276}, E481 (2004).}\BibitemShut {Stop}%
\bibitem [{\citenamefont {Rotter}(2004)}]{McPhase2}%
  \BibitemOpen
  \bibfield  {author} {\bibinfo {author} {\bibfnamefont {M.}~\bibnamefont
  {Rotter}},\ }\bibfield  {title} {\bibinfo {title} {Using mcphase to calculate
  magnetic phase diagrams of rare earth compounds},\ }\href
  {https://doi.org/https://doi.org/10.1016/j.jmmm.2003.12.1394} {\bibfield
  {journal} {\bibinfo  {journal} {J. Magn. Magn. Mater.}\ }\textbf {\bibinfo
  {volume} {272}},\ \bibinfo {pages} {E481} (\bibinfo {year}
  {2004})}\BibitemShut {NoStop}%
\end{thebibliography}%
\end{document}